\documentclass{aa}
\usepackage{natbib}
\usepackage{txfonts}
\usepackage{graphicx}  
\bibpunct{(}{)}{;}{a}{}{,}
\usepackage[english]{babel}
\usepackage{ulem}

\newcommand{\Mwd}{\mbox{$M_\mathrm{wd}$}}

\newcommand{\Rwd}{\mbox{$R_\mathrm{wd}$}}

\newcommand{\Msun}{\mbox{$\mathrm{M_{\odot}}$}}

\newcommand{\Teff}{\mbox{$T_\mathrm{eff}$}}

\def\apgt{\ {\raise-.5ex\hbox{$\buildrel>\over\sim$}}\ }
\def\aplt{\ {\raise-.5ex\hbox{$\buildrel<\over\sim$}}\ }
\newcommand {\be} {\begin{equation}}
\newcommand {\ee} {\end{equation}}

\def\spose#1{\hbox to 0pt{#1\hss}}
\def\simless{\mathrel{\spose{\lower 3pt\hbox{$\mathchar"218$}}
        \raise 2.0pt\hbox{$\mathchar"13C$}}}
\def\simgreat{\mathrel{\spose{\lower 3pt\hbox{$\mathchar"218$}}
        \raise 2.0pt\hbox{$\mathchar"13E$}}}
\def\lta{\mathrel{\spose{\lower 3pt\hbox{$\mathchar"218$}}
        \raise 2.0pt\hbox{$\mathchar"13C$}}}
\def\gta{\mathrel{\spose{\lower 3pt\hbox{$\mathchar"218$}}
        \raise 2.0pt\hbox{$\mathchar"13E$}}}

\begin{document}

\title{Accretion-disc model spectra for dwarf-nova stars}
\author{Irit Idan\inst{1}
          \and
           Jean-Pierre Lasota\inst{2,3}
            \and
              Jean-Marie Hameury\inst{4}
               \and
                Giora Shaviv\inst{1}
} \offprints{J.-P. Lasota, lasota@iap.fr}
\institute{Department of Physics, Technion-Israel Institute of Technology, 32000 Haifa, Israel
             \and
               Institut d'Astrophysique de Paris, UMR 7095 CNRS, UPMC Univ Paris 06, 98bis Bd Arago, 75014 Paris, France
                \and
                 Astronomical Observatory, Jagiellonian University, ul. Orla 171, 30-244 Krak\'ow, Poland
                 \and
                 Observatoire de Strasbourg, CNRS/Universit\'e Louis Pasteur, 11 rue de l'Universit\'e, F-67000 Strasbourg, France
              }

\date{received \today}
\titlerunning{Dwarf-nova disc spectra}
\authorrunning{Idan et al.}

\abstract{Radiation from accretion discs in cataclysmic
variable stars (CVs) provides fundamental information about the properties of these close binary systems and about the physics
of accretion in general. Of particular interest are dwarf-nova
outburst cycles during which variations of the disc properties
allow a detailed study of the physical processes in accretion
flows.} {The detailed diagnostics of accretion disc structure
can be achieved by including in its description all the
relevant heating and cooling physical mechanism, in particular
the convective energy transport that, although dominant at
temperatures $\lta 10^4\,\rm K$, is usually not taken into
account when calculating spectra of accretion discs.
The disc's self-consistently calculated structure and emission
allow testing models of dwarf-nova outbursts and accretion-disc models in general.}
{We constructed a radiative transfer code coupled with a code determining the disc's hydrostatic vertical structure.}
{We have obtained for the first time model spectra of cold, convective accretion discs. As expected, these spectra are mostly flat in the optical wavelengths with no contribution from the UV, which in quiescence must be emitted by the white dwarf. The disc structures obtained with our radiative-transfer code compare well with the solutions of equations used to describe the dwarf-nova outburst cycle according to the thermal-viscous disc instability model thus allowing the two to be combined. For high-temperature radiative discs
our spectra are compatible with models obtained with
Hubeny's code \textsl{TLUSTY}.}
{Our code allows
calculating the spectral evolution of dwarf nova stars through
their whole outburst cycle, providing a new tool for testing
models of accretion discs in cataclysmic variables.
We show that convection plays an important role in determining the vertical disc structure and substantially affects emitted spectra when, as often the case, it is effective at optical depths $\tau\sim 1$. The emergent spectrum is independent of the parameters of the convection model. We confirm that, as required by the disc instability model, quiescent discs in dwarf novae must be optically thick in their outer regions. In general, no emission lines are present in the absence of external irradiation.}

\maketitle

\section{Introduction}
\label{sec:intro}

In most cataclysmic variable stars (CVs) matter lost by the
Roche-lobe filling low-mass secondary star forms an accretion
disc around the white-dwarf primary. In bright systems such as
nova-like stars and dwarf novae in outburst, the disc is the
dominant source of luminosity. In quiescent dwarf novae,
although rather dim, the disc emission provides crucial
information about the physics of accretion discs, in particular
about the mechanisms driving the instabilities that are
responsible for the dwarf-nova outbursts. It is generally
accepted \citep[see, however,][]{sl07} that properties of the
dwarf-nova outburst cycles are reproduced well by the
thermal-viscous disc instability model \citep[DIM; see][for a
review]{l01}, but in reality the description of some phases of
theses cycles is far from satisfactory. This is especially true
of the quiescence state. Until now models of disc emission
spectra have been calculated consistently only for hot, bright
accretion discs \citep[see][and references therein]{wh98}. In
such a scheme the vertical structure of an accretion ring is
calculated using the program TLUSDISK \citet{hubeny90,hubeny91},
whereas the spectra are synthesized by the code SYNSPEC
\citep{synspec}. A different approach to radiative transfer in
accretion discs was presented by \citet[][hereafter SW]{sw91},
but also in this case it was applied  only to hot discs. An
attempt to apply the SW code to cold quiescent discs in
\citet{Idan1} was not conclusive.

The main obstacle in solving
the vertical structure of cold accretion discs is the presence
of convection, which was difficult to include in radiative
transfer codes (Hubeny, private communication); models
of \citet{wh98} are restricted to temperatures for which convection is
unimportant. In their pioneering work \citet{pringletal86} used
vertically integrated disc structures and Kurucz's stellar spectra
to test the DIM versus the model attributing dwarf-nova outbursts
to an increase of mass-transfer rate from the secondary.
More recently \citet[][see also
\citet{nagel2}]{nagel3} have calculated non-LTE
accretion-disc spectra throughout the outburst cycle of the
dwarf nova SS~Cyg, taking into account white-dwarf irradiation.
They found that in order to obtain agreement with observations of
this system in quiescence one has to assume a radial disc
that is optically thin in its outer regions.  Such a structure, however,
is inconsistent with the DIM. The AcDc NLTE code by  \citet{nagel04}
used to calculate these spectra does not take into account
the (often dominant)\,vertical convective energy transport.

Of related interest are radiative-transfer codes describing emission from
cold proto-stellar accretion discs such as \citet{Hugel09} and \citet{min09}. Neither
of these codes includes convection.

There is therefore a need for a
radiative transfer code consistent with the DIM.
This is the main motivation behind the present work. In particular we
have obtained a scheme that is totally consistent with the code of
\citet[][hereafter HMDLH]{hmdl98}, which
allows following the dwarf-nova spectra during
various phases of the outburst with special emphasis put on
quiescence. One should stress that the principal aim of our
project is testing the DIM, not trying to reproduce the observed spectra of dwarf-nova stars at all
costs. In particular
our code should allow calculating various delays in rising to
(and decaying from) outburst between light at different
wavelengths (commonly known a the ``UV-delay") and determining
what properties of the quiescent dwarf-nova disc can be accounted
for by the DIM).

The HMDLH
version of the DIM uses a 1+1\,D scheme in which the
time-dependent radial evolution equations use as input a
pre-calculated grid of hydrostatic vertical structures. These
structures are calculated using the standard equations of
stellar structure (or the grey-atmosphere approximation in the
optically thin case). The angular-momentum viscous transport
mechanism is described by the $\alpha$-ansatz of \citet{ss73}.
For a given $\alpha$, $M/R^3$ and effective temperature $\Teff$
there exists a unique solution describing the disc vertical
structure. Such solutions are very handy for solving the
time-dependent equations of the disc evolution but do not
produce realistic emission spectra. These can be calculated, however, by
using the same input parameters ($\alpha$, $M/R^3$ and $\Teff$)
in a radiative-transfer code on the condition that the vertical
structures calculated by the two methods are the same up to the
photosphere. In this way one can reproduce spectra of the whole
cycle of a dwarf-nova outburst.

The outline of the article is as follows. In Section
\ref{sec:model} we present the model used, with special stress
put on marking the differences with the original SW code on which
it is based. We discuss there also in details the opacities
used. Section \ref{sec:onering} is dealing with the solutions
obtained for various types of physical set-ups. In this section
we compare our vertical structure solutions with those obtained
with the HMDLH code for a hot stationary and cold
non-stationary discs. These two models represent the main
phases of the dwarf-nova outburst cycle respectively: the
outburst and quiescence. We compare the corresponding
thermal-equilibria (``S-curves") in \ref{subsec:sc}. The
importance of convection in determining the vertical disc
structure is discussed in detail in section \ref{subsec:Conv}.
In section \ref{sec:full_disc} we show our hot whole-disc
solutions as examples of spectra emitted during the decay
from maximum of a dwarf-nova outburst. We also show that,
despite differences in treatment of vertical viscosity stratification our solutions
compare very well with the UV spectra obtained by \citet{wh98}.
Then in section \ref{sec:non-stat} we present and discuss the structure and
spectrum of cold quiescent accretion disc. Finally we conclude
the paper in Section \ref{sec:concl} by shortly discussing
future developments and necessary improvements.
A report relating the progress of our work on the radiative-transfer code was published in the conference proceedings \citet{Idan2}. Several solutions presented in this report were only preliminary and have been substantially improved, some results we found to be incorrect.
We address these points in the present article.

\section{The physical model and basic assumptions}
\label{sec:model}

Our model is based on the work of \citet{sw91}. The main
improvements consist in using (modern) line opacities and in
including the convective energy transport. As far as we know our code is the only working accretion-disc radiative-transfer programme which consistently includes convection. The last improvement allows combining the radiative transfer programme with the DIM code of HMDLH where the time-dependent evolution of dwarf novae is calculated.

\subsection{Assumptions}

We describe an accretion disc as composed of concentric rings
orbiting the central gravitating body of mass $M$ at a
Keplerian angular speed
\be
\Omega_{\rm
K}=\sqrt{\frac{GM}{R^3}}.
\label{eq:kepler}
\ee

The disc is assumed to be geometrically thin, i.e.
\begin{equation}
\label{eq:geo_thin}
z_0 <<R,
\end{equation}
where $z_0$ is the disc height. This assumption allows
decomposing the disc equations into their vertical and radial
components. In this paper we will solve only the vertical
accretion-disc equations and obtain the corresponding emission
spectra. However, since the radiation of a steady-state
accretion disc can be considered as a sum of radiation from
individual rings our results can directly applied to such
configurations. The same is true for quasi-stationary phases
of dwarf-nova outbursts. Also radiation from non-steady quiescent
discs of dwarf-nova stars can be treated as being the sum of the
emission of individual rings. This is true as long as radial
gradients are much smaller than the vertical ones.
During the outburst of dwarf nova, when the temperature and
density fronts are propagating through the disc this assumption
is no longer generally valid. However, since radial gradients are
comparable to vertical gradients only in a narrow zone close to the
fronts the error on integrated spectra cannot be very large

The disc is assumed to be in a vertical hydrostatic equilibrium and in LTE.

\subsection{Equations}

The vertical structure of a ring at a distance $R$ from the
centre is found by solving the following equations:

\begin{itemize}

\item the hydrostatic equilibrium equation
\begin{equation}
\frac{dP}{dz}=-\rho\,g_z=- \rho\,\Omega^2_{\rm K}\,z,
\end{equation}
where $P$ is the total (gas + radiation) pressure;

\item the energy balance equation
\be
\frac{dF_{\rm z}}{dz} = Q_{\rm vis} =  \left(\frac{dF_{\rm
z}}{dz}\right)_{\rm rad} + \left(\frac{dF_{\rm z}}{dz
}\right)_{\rm conv},
\ee
where $F_{\rm z}$ is the energy flux in the vertical ($z$) direction.
We use the $\alpha$ prescription of \citet{ss73} to describe
the viscous energy generation
\begin{equation}
Q_{\rm vis}=\frac{3}{2} \alpha \Omega_{\rm K}P.
\label{qvis}
\end{equation}
Equation (\ref{qvis}) defines therefore the stratification
of the viscous energy generation;

\item the radiative energy flux is given by
\begin{equation}
\label{eq:temp}
\int_0^\infty\,\left(J(z,R,\lambda)-B(T(z,R),\lambda)\right)\kappa(\lambda) d\lambda=Q_{\rm vis} -
\left(dF/dz\right)_{\rm conv},
\end{equation}
where $J$ is the mean intensity, $B$ the Planck function and $\kappa(\lambda)$ is the
monochromatic mass absorption coefficient;

\item the radiative transfer equation is
\begin{equation}
\mathbf{n} \cdot \mathbf{\overrightarrow{\nabla}} I=\kappa(S-I),
\end{equation}
where $I$ is the intensity, $S$ the source function and $\mathbf{n}$ is a unit vector in the ray direction.
This equation is solved in the two-stream approximation (see \ref{subs:method}).
\\
\item As in HMDLH, the convective energy flux is calculated by the
    method of \citet{p69}. Whenever the
    radiative gradient
\be
\nabla_{\rm rad}\equiv \left(\frac{d\ln T}{d \ln P}\right)_{\rm rad}
\ee
is superadiabatic, the temperature gradient of the
structure $\nabla$ is convective ($\nabla=\nabla_{\rm
conv}$). The convective gradient is calculated in the
mixing length approximation, with the mixing length taken
as $H_{\rm ml} = \alpha_{\rm ml} H_P$, where $H_P$ is the
pressure scale height:
\begin{equation}
H_P = {P \over \rho g_{\rm z} +(P\rho)^{1/2} \Omega_{\rm K}},
\end{equation}
which ensures that $H_P$ is smaller than the vertical scale height of the
disc.
The convective gradient is found from the relation:
\begin{equation}
\nabla_{\rm conv} = \nabla_{\rm ad} + (\nabla_{\rm rad} - \nabla_{\rm ad}) Y
(Y+A),
\end{equation}
where $\nabla_{\rm ad}$ is the adiabatic gradient, and $Y$
the solution of the cubic equation:
\begin{equation}
{9 \over 4} {\tau_{\rm ml}^2 \over 3 +\tau_{\rm ml}^2} Y^3 + VY^2 + V^2 Y -V
= 0,
\end{equation}
where $\tau_{\rm ml} = \kappa \rho H_{\rm ml}$ is the optical depth of
the convective eddies. The coefficient $V$ is given by:
\begin{eqnarray}
V^{-2} = \left( {3 + \tau_{\rm ml}^2 \over 3 \tau_{\rm ml}}\right)^2 {g_{\rm
z}^2 H_{\rm ml}^2 \rho^2 C_P^2 \over 512 \sigma^2 T^6 H_P} \left( {\partial
\ln \rho \over \partial \ln T} \right)_P \times (\nabla_{\rm rad} - \nabla_{\rm ad}).\nonumber \\
 \,
\end{eqnarray}
As in HMDLH in our model we take $\alpha_{\rm ml} = 1.5$.
\\

Here we should mention a caveat. The description of
convection used in the present version of the code is not
necessarily the correct one as it is not even clear that
mixing-length approximation can be applied to turbulent
accretion discs. However, in the context of the
``$\alpha$--disc paradigm" this is probably the best that
can be done. It is an efficient method of describing an
accretion disc in which most of the vertical energy
transport is not radiative. Finally, as mentioned above,
this treatment of convection is used in the HMDLH code
which is the main reason for using it also in the present
programme. If this approximation is applicable
to accretion discs, there are no indications as to what
the value of $\alpha_{\rm ml}$ should be. In solar models values between 1 and 2 are used
\citep[see][and reference therein]{hmdl98}; in \ref{subsec:Conv} we discuss the effect of various values $\alpha_{\rm ml}$ on the vertical
structure.

\item The outer boundary condition is
\be
D(r)=\int_{0}^{Z_0}Q_{\rm vis} dz=\sigma T_{\rm eff}^4.
\ee

In the case of a stationary disc
\begin{equation}
\label{1}
\dot{M} = 2 \pi R \int_{0}^{Z_0}\rho v_r dz =constant,
\end{equation}
where $v_r$ is the radial flow velocity
one has
\begin{equation}
D(r)=\frac{3}{8\pi}\dot{M}\frac{GM_{\rm wd}}{R^3}
\left(1-\left(\frac{R_{\rm wd}}{R}\right)^{1/2}\right),
\end{equation}
but our solutions apply to any distribution of accretion rate $\dot M(r)$ in
a geometrically thin accretion disc.
\end{itemize}

\subsection{Input (opacities and  EOS)}
\label{subs:opacities}

Several features of the original SW code have been improved but the use of modern opacities and especially of the line
opacities is a major modification we present here in some
detail.

There are several ways of calculating opacities. One can either use codes such as PHOENIX \citep{phoenix} or ``Atlas 12"
\citep{kurucz}, or one can use the OPAL database. Our updated
version of the SW code can use the opacities from ``Atlas 12".
However, since calculating the opacities is time--consuming we
decided to use tabulated data from the Opacity Project (OP) database. To
tabulate the data we used the data from the Opacity Project -
$OPCD\_3.3$ that was downloaded from the website at the Centre
des Donn\'{e}s de Strasbourg (CDS). For a given abundance
mixture the subroutines $mx.f$, $mixv.f$, $opfit.f$ and
$mixz.f$ allowed us (with some modifications) getting the
tabulated opacities  for the wavelengths chosen as our basic
grid. For the emerging spectra from every ring we use 10,000
wavelengths starting from $100${\AA} up to $10^6${\AA}.  The
grid interval varies between $\Delta \lambda =0.25${\AA} for
the most important zone such as the range between
$850-2300${\AA} and $1-2${\AA} used for wavelengths shorter
than $800${\AA} or above $2300${\AA}. For wavelengths longer
than $5000${\AA}, $\Delta \lambda$ varies logarithmically
starting at $2${\AA} and getting up to $200${\AA} for the very
long wavelengths. All data were tabulated using a grid for
the temperature $T$ and electron density $N_e$. The indices
$ite$ and $jne$ are defined by
\begin{equation}
    ite=40x\log(T),  \ \ jne=4x\log(Ne),  \ \
    \triangle(ite)=\triangle(jne)=2
\end{equation}

The mesh in the tables is calculated for $140<ite<320$
therefore the lowest temperature that can be used in the code
is $3160 \rm K$. Let us stress that the opacities in the
present work do not take into account the global line
broadening due effects such as velocities, expansion
opacity \citep[see][]{sw05} or microscopic effects such as microturbulence.
Consequently, the lines provide only an indication of the presence of a given ion and
cannot be used, so far, for abundance determinations etc. These broadenings will be
included in the code in the near future.

\begin{figure}
\centering
 \vspace*{1em}
 \includegraphics[width=0.9\columnwidth]{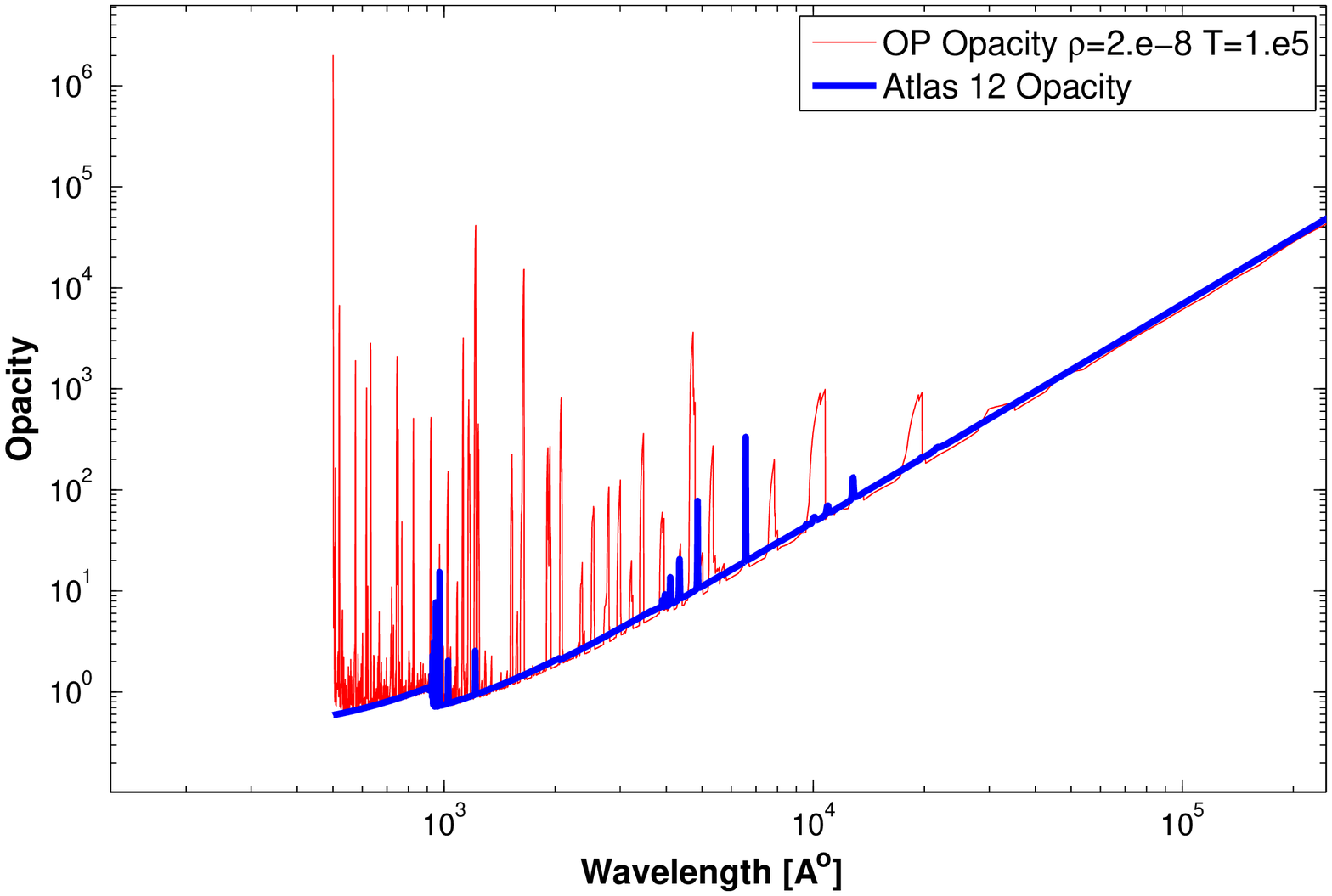} \\
 \centering
 \includegraphics[width=0.9\columnwidth]{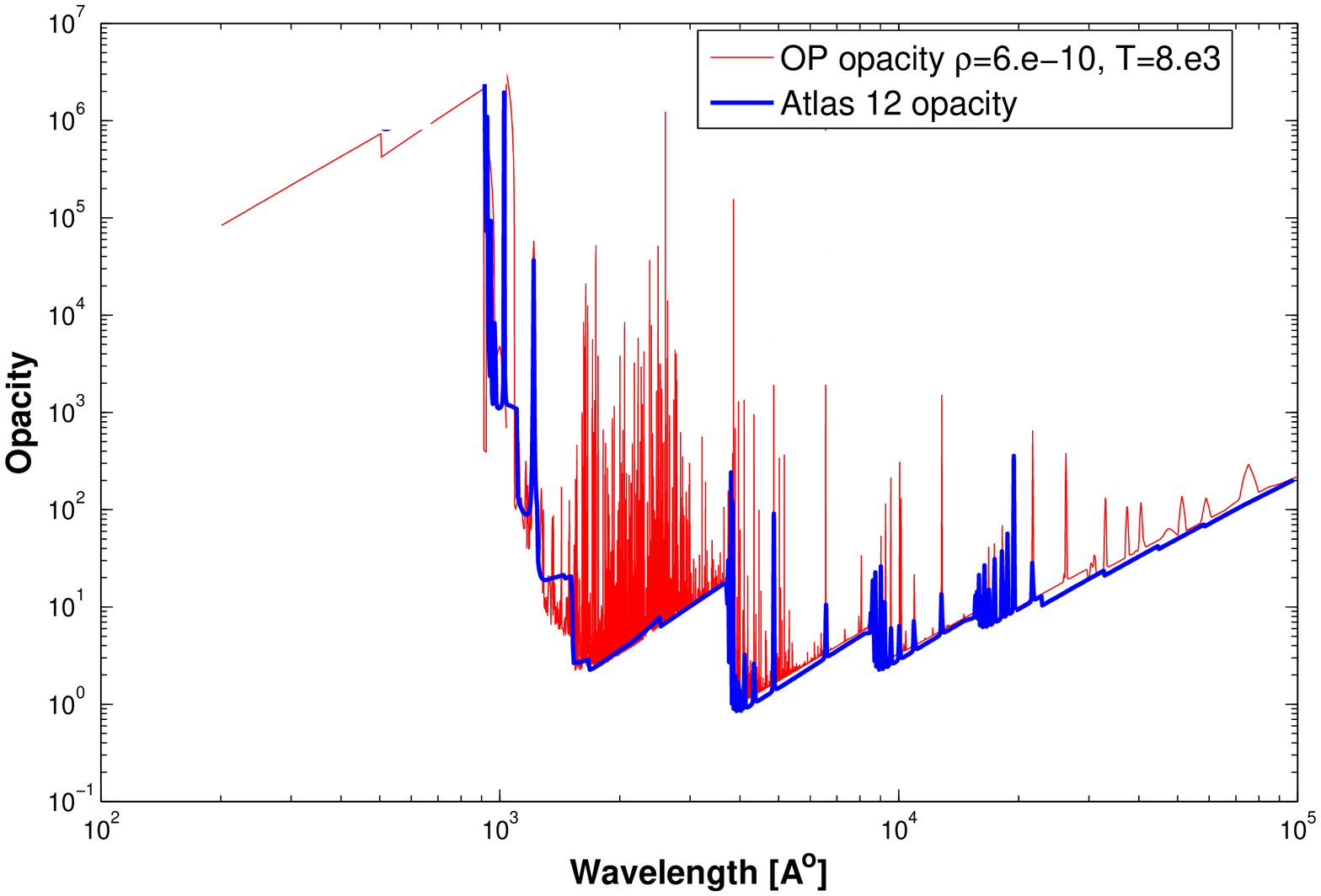}
 \caption{Comparison between the opacities as function of wavelength obtained
  from ``Atlas 12" of Kurucz and those obtain from OP for $T=10^5\,\rm K$ and
  $\rho=2\times10^{-8}\rm g\,cm^{-3}$ (upper figure)
  and $T=8\times 10^3\,\rm K$ and $\rho=6\times 10^{-10}\rm g\,cm^{-3}$ (lower figure).
  The agreement between the two is more than satisfactory. {(OP opacities are for solar
  abundance mixture while the Atlas ones contain just continuum plus H and He lines.)}}
 \label{fig:opacity}
\end{figure}

The reason for allowing the code to work either with OP tabulated data or with ``Atlas 12" was to test the tabulated opacities. In Figure
\ref{fig:opacity} we show the comparison of opacities as function of wavelength calculated for $T=10^5$K and
$\rho=2\times10^{-8}$ g cm$^{-3}$, and $T=8\times10^3$K and
$\rho=6\times 10^{-10}$g cm$^{-3}$ using either ``Atlas 12" or the OP. The ``Atlas 12" continuum opacities used were for a solar mixture with the
addition of only  H and He lines. The opacities from OP are also for a mixture of solar abundance (which includes the lines and the
continuum). The result of test shows that tabulated OP opacities are reliable.

The two stream approximation contains the scattering coefficient.
Following the good agreement obtained between ``Atlas 12" and OP, the scattering
coefficient were calculated from ``Atlas 12" taking into account scattering from H, He and electrons. One should stress, however, that scattering is
not important in CV discs.

The Rosseland mean opacities are taken from OPAL tables for
solar composition. While solving the radiative transfer
equation we checked for the consistency between the Rosseland
mean opacities obtained from the solution of the radiative
transfer equation and the values from the OPAL table. The
equation of state  is interpolated from the tables of
\cite{FGV}.  The computer programme presented in this
article is able to produce accretion disc models for
temperatures from 3160 to 100 000 K\footnote{However,
we do not attempt to produce \textsl{stationary} accretion
discs spanning such a range of temperatures since such discs do \textsl{not} exist.}.
Consistently with the lower temperature limit we not take into account molecular processes and opacities.
Rayleigh scattering is also neglected. Most of the
reported quiescent dwarf-nova disc effective (colour)
temperatures ($\gta 3000K$) justify this assumptions.

\subsection{The Solution Method}
\label{subs:method}

The basic idea behind the SW original code was to couple the
hydrostatic and the radiative transfer equations. To achieve
this the programme iterates the two equations at the same time.
The iteration is repeated until a structure is found to which
both the solutions of the hydrostatic and the radiative
transfer equations converge.

The temperature profile is determined from Eq. (\ref{eq:temp}).
The temperature $T(z,R)$ in this equation is solved by
expressing $J$ in terms of $B$ using the radiative transfer
equation \citep[for a detailed description see][]{w81,kw84}. As
mentioned above we assumed that each radial ring is independent
of its neighbours rings so we consider a vertical radiative flux
only. The radiative transfer equation is solved in the two
stream approximation. $I^+$ represent the specific
intensity in the outward direction and $I^-$ is the specific
intensity in the inward direction. The radiative equation is
therefore written as:
\begin{eqnarray}
\pm\frac{dI^{\pm}(R,z,\lambda)}{dz} &=& -(\kappa(\lambda)+\sigma(\lambda))I^\pm(R,z,\lambda)
\nonumber\\
&  & +\sigma(\lambda)J(R,z,\lambda)+\kappa(\lambda)B(T(R,z),\lambda)
\end{eqnarray}
where $\sigma(\lambda)$ is the scattering coefficient. The
number $N$ of grid points is the same in the radiative transfer
and the hydrostatic calculations. The boundary conditions are:
(1) no incident flux ($I^{-}_{j,1}=0$), (2) at the symmetry plane
the upward and downward specific intensities  are equal:
$I^{+}_{j,N}=I^{-}_{j,N}$ where j designates the ring number.

As is now well known, when solving the vertical-structure
equations of an accretion disc the main problem is that
contrary to the case of stellar atmospheres the exact height of
the photosphere is unknown a-priori. Because of the gravity
increasing with height, a change in the position of the
photosphere is followed by a change of the gravitational
acceleration which leads to a change in the entire vertical
structure of the disc. One has to remember that the upper part
of the disc is optically thin and (depending on the
assumptions) may be the site of a non negligible energy
production. However, since this optically thin region is a poor
radiator and absorber, even a small energy production in this
region may have large effect on the final temperature
structure.

In view of the above situation we approached the problem in the
following way:  we guessed first an initial
temperature--optical depth relation and assumed an initial
height $z_0$.  The optical depth (measured vertically from
outside the disc towards the midplane) at the disc's uppermost
point was chosen to be between $10^{-2}-10^{-6}$,
and the validity of this assumption was checked once the final
structure was obtained. Using the T($\tau$) relation we can
integrate the hydrostatic equation from $z_0$ down to $z=0$.
One should emphasize that $\tau$ in the current paper is
calculated using the Rosseland mean opacity and not the optical
depth at 5000{\AA} as in SW. When the height $z_0$ is the
correct height the total energy generation in the ring
calculated from the energy equation is equal to the total
luminosity of that particular ring. If this were not the case
we were altering the value of $z_0$, keeping the T($\tau$)
relation fixed, until the energy condition was satisfied. Once
this iteration converged the hydrostatic model was
consistent with the energy generation but not yet with the
radiation field.
\begin{figure}
 \centerline{\includegraphics{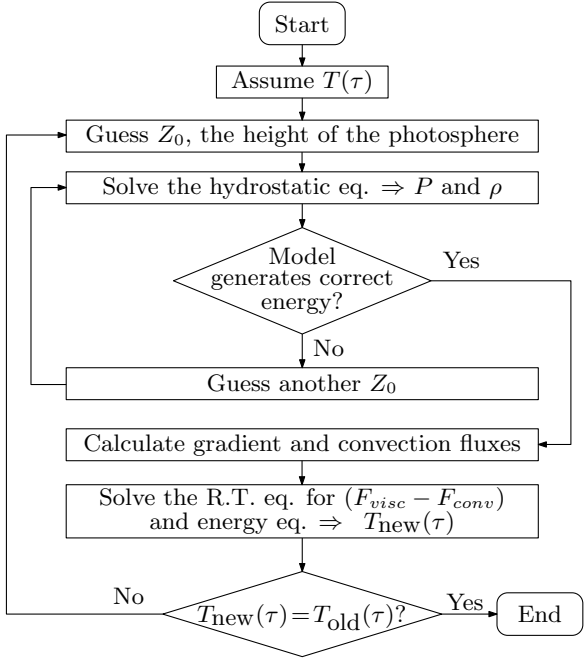}}
 \caption{The iteration scheme.}
 \label{fig:iteration_sch}
\end{figure}

We then turned to solving the radiative transfer equation using the disc height, the pressure and the density structure
obtained from the hydrostatic iteration. This structure was
kept fixed during the iteration for the radiative field.
The first step in solving the radiative transfer equation is to calculate the convection.
For each grid point we calculated the convective flux.
If the convective flux was dominant (more than $80\%$ of the viscous flux) - the temperature was determined by the adiabatic gradient.
In the opposite case the convective flux was subtracted from
the total (viscous) flux and
the solution of the radiative transfer equation for the remaining flux, yielded
a new $T(\tau)$ relation. The temperature profile was assigned
new values for the old values of optical depth. If the new
$T(\tau)$ profile from the radiative iteration agreed (to a
chosen degree of accuracy; usually $0.5\%$) with the old $T(\tau)$ relation from the hydrostatic iteration, the hydrostatic structure was
consistent with the radiation field. If not, we iterated again the hydrostatic equation
using the new $T(\tau)$ relation.
\begin{figure}
 \vspace*{1em}
 \includegraphics[width=0.9\columnwidth]{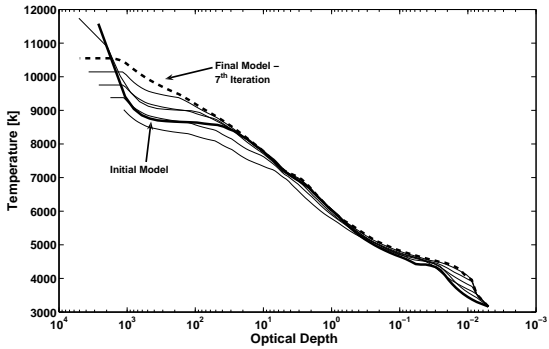}
 \caption{Variation of the temperature-optical depth profile during the iteration to the converged
  solution ($\Mwd=0.6, \Msun$, $R=5\Rwd$, $\alpha=0.03$ $\Teff=5000$K). The inner part of the structure
  is convectively dominated up to $\tau=3 $.
  In this case the solution was obtained after seven iterations. In the last iteration the temperature
  profiles obtained from vertical equilibrium and radiation field differ by less than $0.5\,\%$}
 \label{fig:iterationt}
\end{figure}

In Figure \ref{fig:iterationt} we show how the temperature
profile changes during the iteration for a cold accretion disc.
In this case the convergence was obtained after $7$ iterations.
One has to remember that since the location of the photosphere
is unknown  at the beginning of the iteration we must iterate
for it so as to satisfy the energy conservation condition.
The initial temperature optical-depth profile is a priori unknown
and we guess a certain $T(\tau)$ law. This $T(\tau)$ might change
dramatically after the first iteration but afterward quickly converges to the correct solution.


To summarize we have four basic iteration loops in the code
(see Fig. \ref{fig:iteration_sch}):
\vspace*{-0.5em}
\begin{enumerate}
  \item Calculating the hydrostatic structure using a given $T(\tau)$ relation.
  The output results are disc height, pressure and density structure.
  \item Calculating the convective flux using the structure of the hydrostatic structure and subtracting it from the total flux.
  \item Solving the radiative transfer for a given disc height, $P$ and $\rho$.
  The output result is a new $T(\tau)$ relation.
  \item Repeating the previous two iteration until the $T(\tau)$ relation
  from the hydrostatic part is in agreement with the $T(\tau)$ relation from the radiative
  transfer equation.
\end{enumerate}
Only after those iterations have converged the flux emerging  from the disc is  calculated using the radiative transfer equation.The vertical structure
of the ring itself is divided into 100 $z$ points  (100 layers) and we use 10000 wavelengths to calculate the radiative transfer equation and the
emerging spectrum.

Once lines opacities are included in the code the
calculation of the spectrum must be done with a much finer
grid. The contribution of the lines to the total energy budget
in the disc when solving the radiative transfer equation is
negligible. Therefore 100 grid points can be used to obtain the convergence of the hydrostatic and radiative transfer
equations. However, for the calculation of the emergent
spectrum we interpolate between the 100 vertical grid points to up to 500 points in order to get better accuracy. In a recent progress report \citep{Idan2} we announced the presence of emission lines in the spectrum, unfortunately this was an artefact of using only 100 grid points in the calculation. Once a finer grid was used, the alleged lines disappeared.
This just indicates the importance of the accuracy in calculations of the spectra.

The entire disc is divided into a series of concentric rings
and their width is determined according to their distance from
the WD. All rings which are  further away than  $2\Rwd$ are
assumed to have a  width of $1\Rwd$, below $2\Rwd$ the width of the ring is taken to be $0.05\Rwd$, which allow us a detailed
calculation of  the emission of the rings near the boundary
layer.

The above procedure is repeated for each ring. The emerging fluxes from the rings are then integrated  in order to get the total flux from the entire
disc according to:
\begin{equation}
\label{totflux}
    F_{\rm total}=2\pi \int_{R_{\rm in}}^{R_{\rm out}} F(R)R\,dR
\end{equation}

Since the radiative transfer equation is solved in the
two-stream approximation the resulting spectrum does not
include the effect of limb darkening. The angle dependence
will be added to our code in a future paper through a SYNSPEC-type
code. In the present article, however, we will show spectra
convolved with an instrumental profile in the manner of the
ROTIN\footnote{http://nova.astro.umd.edu/Synspec43/synspec-frames-rotin.html}
programme.

\subsection{The disc instability model}
\label{subsec:dim}

Since the main objective of our work is obtaining a tool allowing tests of the dwarf-nova Disc Instability Model \citep[DIM; see][for a review]{l01} we will briefly recall its main tenets that are relevant to the present work.

In the DIM the dwarf-nova outbursts are attributed to a thermal-viscous instability appearing when the
effective temperatures are contained between $5000 \lta \Teff \lta 7000$K (these critical temperatures depend
only very weakly on the other disc parameters). Therefore stationary discs with effective temperatures
in this range do not exist in the Universe. When the rate at which a disc is fed with matter corresponds to
an unstable configuration the result is an outburst cycle during which the disc oscillates between the hot and cold
(quiescent) states. Outbursts are triggered by heating fronts propagating up (inside-out) or down (outside-in)
the surface-density gradients. During the rise of the outburst the disc is hot downstream and cold upstream
of the front. Finally, at maximum most of the disc is brought to
a hot quasi-stationary state with the characteristic $\Teff\sim R^{-3/4}$ profile.
A cooling front (always outside-in) brings the disc back to the cold (quiescent) state through
a sequence of quasi-stationary configurations with decreasing accretion rates. In quiescence
the effective temperature is everywhere lower than characteristic temperature (Eq. \ref{eq:Teff_crit_low}) and is roughly constant
with radius \citep[as conformed by observations, see e.g.][]{froning99}.
On the other hand the surface-density and the accretion rate strongly increase with radius (in quiescence the disc
is filling up). The fact that spectra of quiescent dwarf-novae
cannot be fitted with constant accretion-rate disc models is confirmation of this prediction of the model.
All the phases of the outburst cycle have characteristic spectral signatures.

Locally (for a given ring), for a given set of parameters (white-dwarf mass, viscosity parameter) the disc equilibria form
an S-shaped curve on the $\Sigma-\Teff$ plane (see Fig. \ref{fig:scurve}). During the decay from maximum the surface
density and effective temperature of the hot parts of the disc decrease along the upper branch of the S-curve until
they reach the critical value below which no stationary solution exists. The value of the critical surface-density
\begin{equation}
\Sigma^+_{\rm crit}= 48.9 \left(\frac{\alpha}{0.1}\right)^{-0.77}\left(\frac{M}{\Msun}\right)^{-0.37}
\left(\frac{R}{10^{10}}\right)^{1.11} \rm g\,cm^{-2}
\label{eq:sigmamin}
\end{equation}
implies that the outer disc in quiescence must be optically thick; especially that after leaving the
upper branch its surface density {\sl increases} before it settles on the lower (cold) branch of the
S-curve. Therefore observations and spectral models that requires an optically thin quiescent outer disc in
dwarf-nova stars contradict the DIM (see Sect. \ref{sec:non-stat}).

\section{Single-ring calculations}
\label{sec:onering}


\subsection{The S-curves}
\label{subsec:sc}

Since we intend to calculate spectra of disc states given
by the time dependent code of HMDLH we checked that
the vertical structures calculated by the two codes are
sufficiently close for the whole procedure to make sense.
We compared in detail hot and cold disc vertical structures.

The principal difference between the two calculated disc structures
is their heights. In the HMDLH case the boundary condition is fixed at $\tau=2/3$ whereas in ILHS it is given by $\tau\approx 0$
($\sim 10^{-2}-10^{-6}$).

\begin{figure}
 \vspace*{1em}
 \includegraphics[width=1.1\columnwidth]{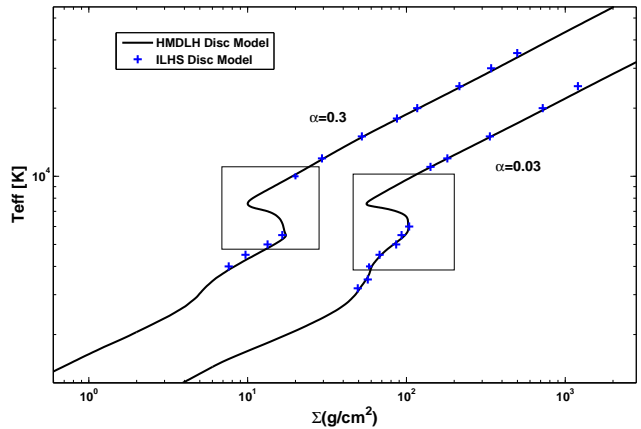}\\
 \caption{$\Sigma-\Teff$ curves at $R=5\Rwd$ with $\Mwd=0.6\Msun$,
  for $\alpha=0.3$ (left) and $0.3$ (right). In solid line the curved obtain from HMDLH model
 and in ``$+$" the results obtained with ILHS. The boxes represent the zones where
 convection dominates.}
 \label{fig:scurve}
\end{figure}

The best way of comparing the two structure is by plotting the thermal equilibria obtained by the two methods.
As mentioned above (Sect. \ref{subsec:dim}), in the $(\Sigma,\Teff)$ plane they form an S-shaped curve. Figure \ref{fig:scurve} shows
an example of $\Teff(\Sigma)$ S-curves obtained by the HMDLH and ILHS codes for $\Mwd= 0.6\,\Msun$ at a radius of $5\Rwd$ \citep[we used the white-dwarf mass-radius relation of][]{nau72}. Each point on the S-curve represents a thermal-equilibrium state of the accretion disc at that radius. The upper branch of the S-curve corresponds to hot stable solutions, while the lower branch where the temperature is below $\simeq 6000\,K$, represents cold, stable configurations. The middle branch of the S-curve contains thermally unstable equilibria.

The agreement between the results of the two codes is very good on both branches of the S-curve. On top of the changes in the
opacities due to the hydrogen ionization, the disc close to the
turning knees of the S-curve (the unstable zone) is fully
convective (see Fig. \ref{fig:scurve}). We were unable to calculate the unstable branch
with our radiative transfer equation code. This is of not much
importance as the equilibrium states represented by the middle
branch are physically meaningless. However, we were also unable to obtain the convection-dominated disc vertical structures between $\Teff \approx 10^4\,\rm K$ and the critical hot-branch temperature $\approx 7000\,\rm K$ since not enough radiative flux was left to use in the transfer code. This is a more serious drawback which somewhat limits the code's application. We are working on trying to fix this problem.

On the hot branch, for $\Teff \approx 10^4\,\rm K$ the convective flux transports $\sim 100\%$ of the flux.
Configurations with temperatures down to $\Teff\approx 4000\,K$ are also convection dominated. For example at $R=5\Rwd$, for 5000\,K 70\,\% of the disc vertical extent is convective and convection still transports $\sim 100\,\%$ of the flux. For effective temperature of 4000\,K 35\,\% of the disc has a
convective flux which represent 60\,\% of the total flux.

The difference between the values of $\Sigma_{\rm max}$ (the
maximum surface density on the lower branch) calculated by the
two codes is about $2\,\%$.

The excellent agreement between the low-temperature solutions calculated by the two methods
shows that molecular opacities do not affect the disc structure: they are taken into account in
HMDLH, but not in our code.

\subsection{The effect of convection}
\label{subsec:Conv}

The discussion in the preceding subsection suggest that the effects of convection are worth a closer look.

\begin{figure}
 \vspace*{1em}
 \includegraphics[width=0.9\columnwidth]{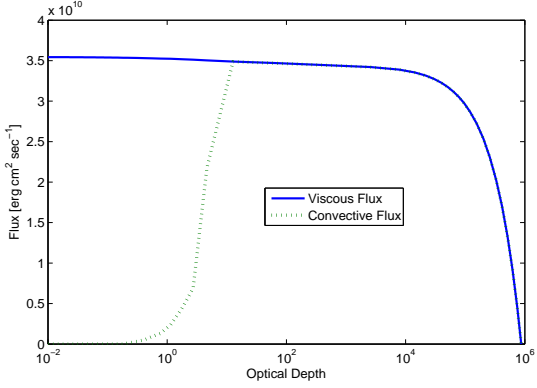} \\
 \caption{The total (viscous) flux and the convective flux (dashed line) as function of the  total optical
 depth  calculated for a cold disc with $\alpha=0.03,\ \Teff=5000\,K, \Mwd=0.6\Msun$, at $R=75R_{wd}$}
 \label{fig:fconv}
\end{figure}

\begin{figure}
 \vspace*{1em}
 \centering
 \includegraphics[width=0.9\columnwidth]{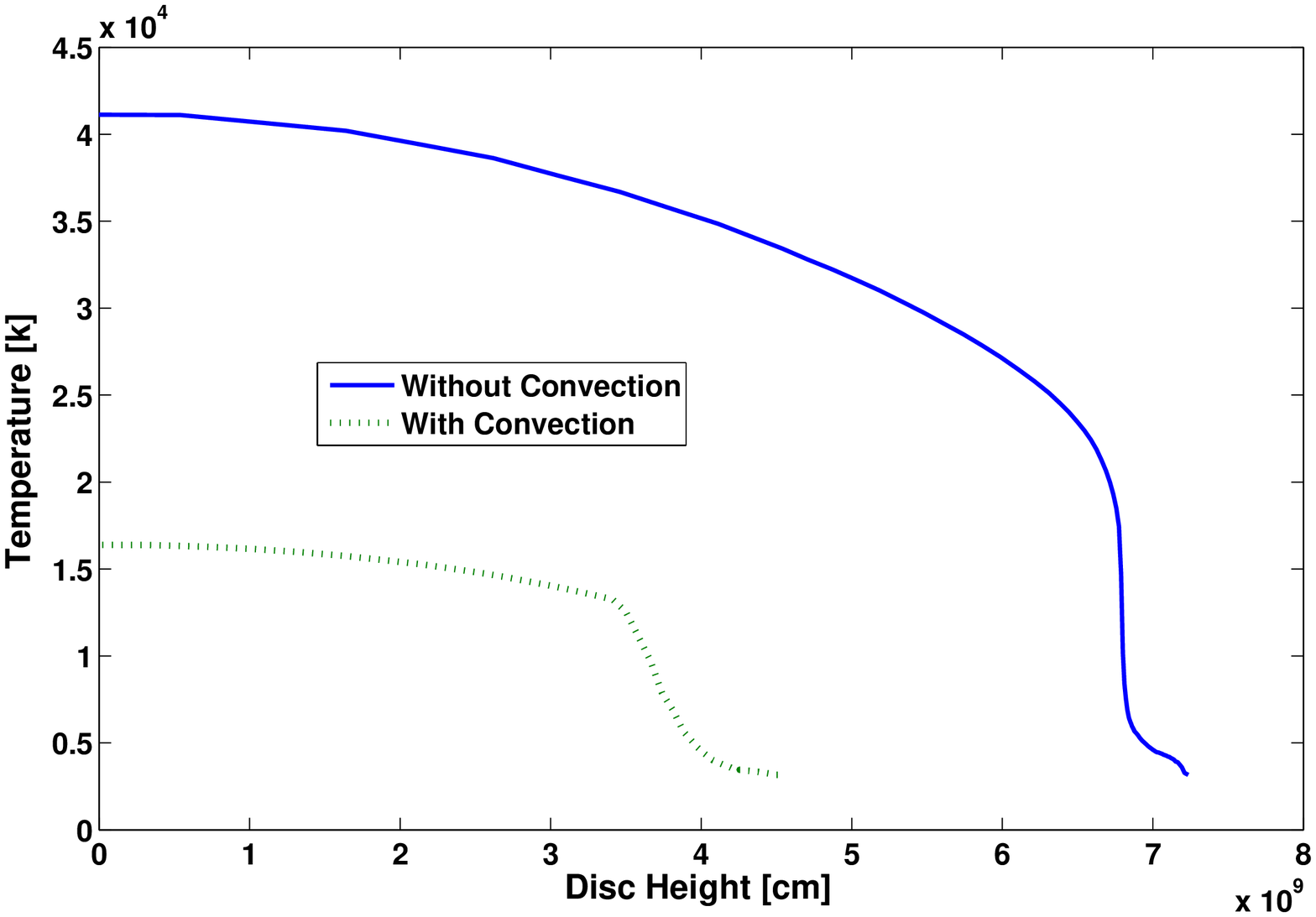} \\
 \includegraphics[width=0.9\columnwidth]{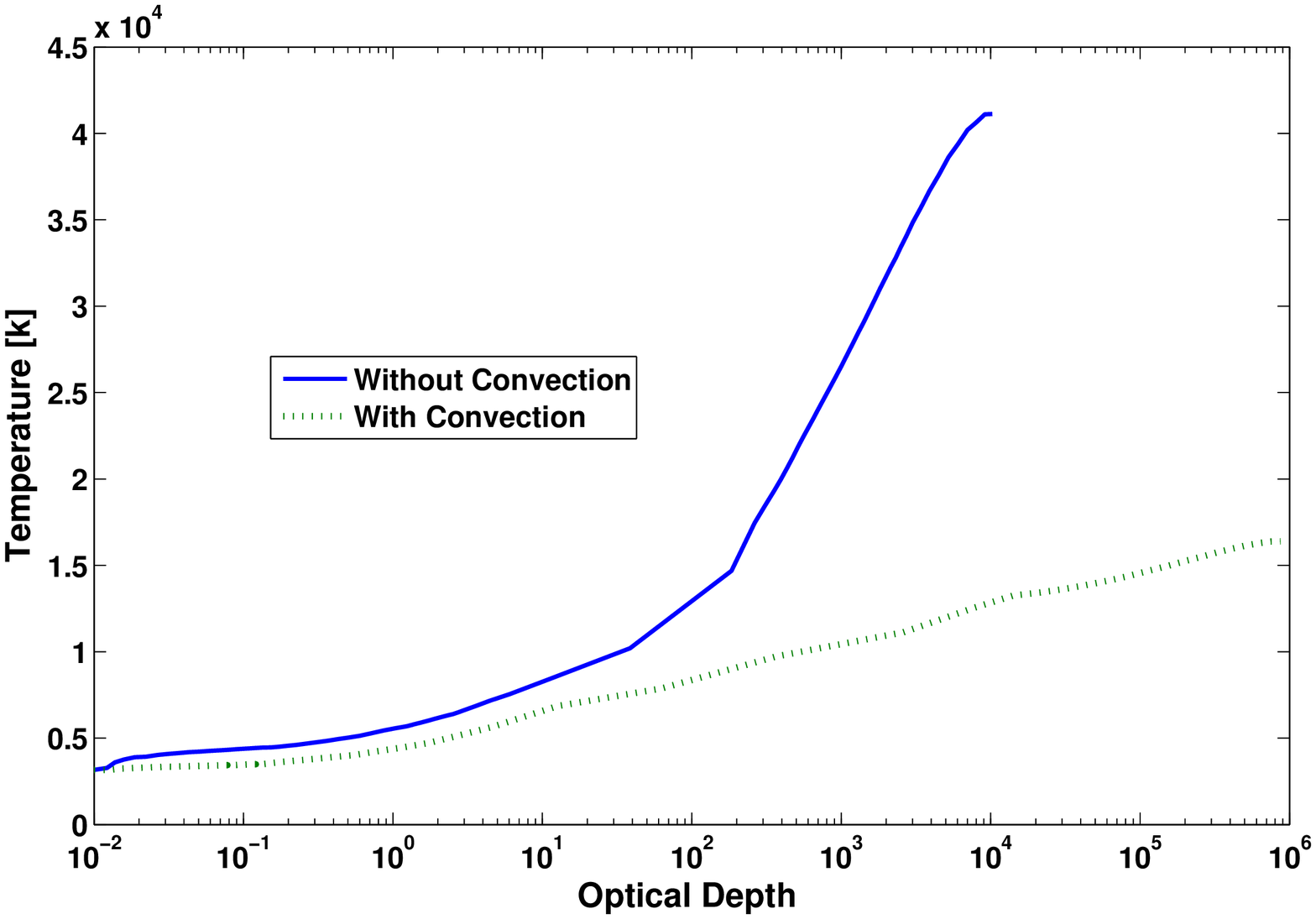} \\
 \caption{Comparison between the temperature profiles for models with and without
 convection for cold disc at
 R=75$\Rwd$ and $\Teff$=$5000$K and
 for $\alpha=0.03$.}
  \label{fig:conv_eff}
\end{figure}

At temperatures corresponding to hydrogen recombination convection begins to play an important role in the vertical energy transport. Therefore at these and lower temperatures it must be taken into account when determining the vertical structure of accretion discs. In Fig. \ref{fig:fconv} and Fig. \ref{fig:conv_eff} we present the vertical structure of a disc ring with effective temperature of $5000$\,K and $\alpha=0.03$ at a distance of $75\Rwd$. Convection dominates the inner (close to midplane) part of the disc up to $\tau\approx 10$. (At $\Teff \approx 10^4\,\rm K$ the convection zone reaches  $\tau\approx 2$).

Figure \ref{fig:conv_eff} exhibits the influence of convection on the temperature profile. Once convection is included the vertical structure becomes cooler and denser. Therefore the disc is becoming optically thicker - for the model shown here the optical depth changes from $\tau \approx 10^4$ without convection, to almost $\tau \approx 10^6$, when the latter is included.

\begin{figure}
 \vspace*{1em}
 \includegraphics[width=0.9\columnwidth]{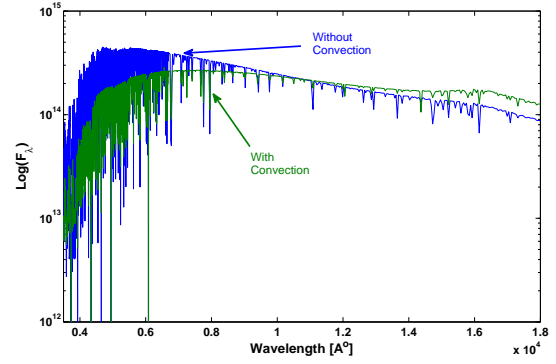} \\
 \caption{Comparison between the fluxes for models with and without convection for a cold disc at
 R=75$\Rwd$ and $\Teff$=$5000$K and
 for $\alpha=0.03$. Convection is still efficient for $\tau\gta 1$ which affects the emerging spectrum.}
 \label{fig:flux_conv}
\end{figure}

Therefore effects of convection noticeably affect the emerging spectrum. According to Figure  \ref{fig:conv_eff} at $\tau=1$ the temperature of a convection dominated disc is 1000K cooler than a radiative disc with the same effective temperature. The different temperature structure implies therefore different emergent spectra. Figure \ref{fig:flux_conv} presents the differences between the corresponding disc spectra obtained with and without convection.

The calculations presented in this paper do not include disc irradiation by the white dwarf and boundary layer. In principle, by heating up the disc surface irradiation may reduce the effects of convection \citep[see e.g.][]{hld99}. However, in practice, only the innermost regions of a
quiescent dwarf-nova disc will be strongly modified by irradiation \citep{hld99}, the rest of the disc being only weakly modified especially that the optical depth increases with radius. There irradiation might affect the spectrum by producing emission lines \citep{Smak91}.

\begin{figure}
 \vspace*{1em}
 \includegraphics[width=0.9\columnwidth]{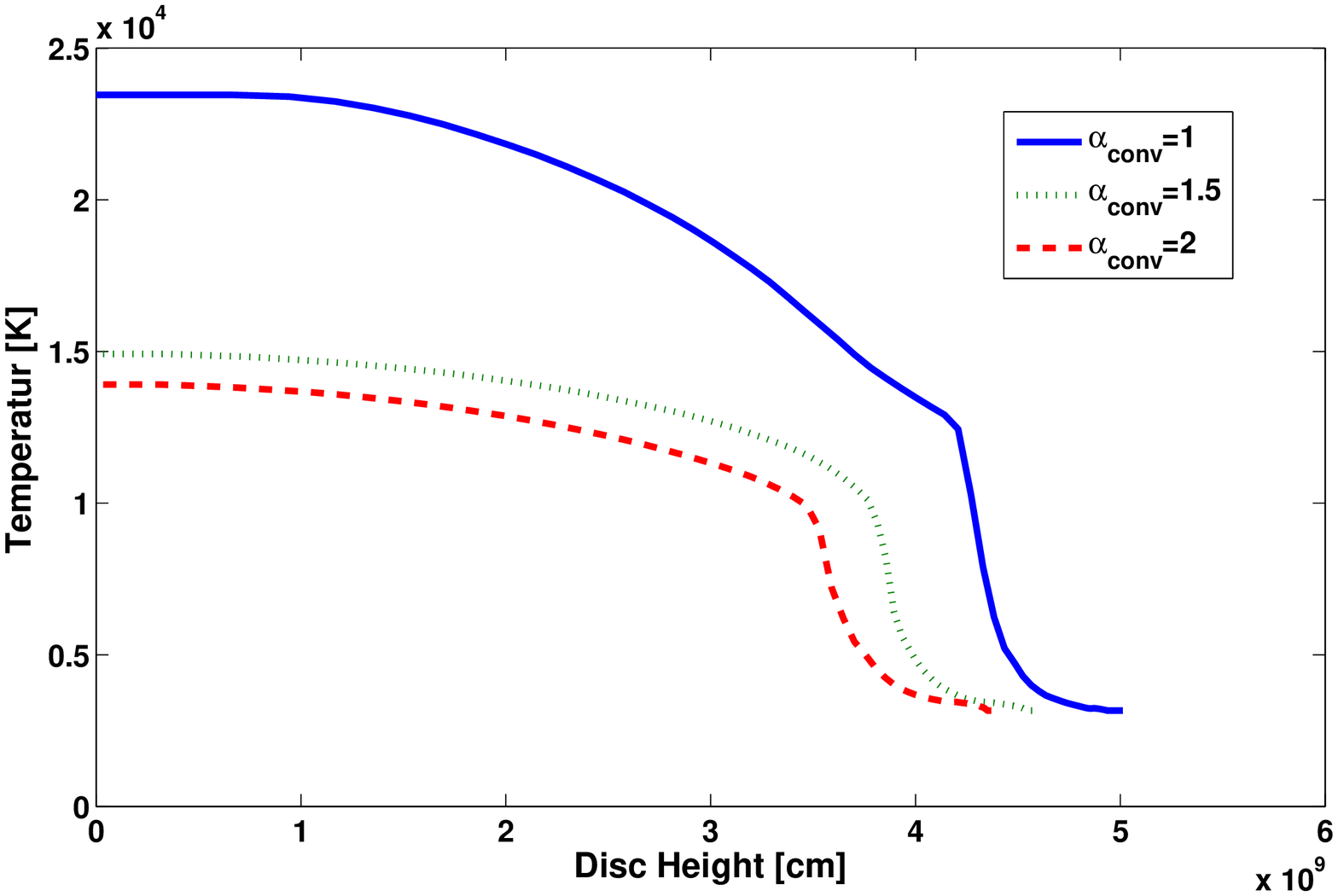} \\
 \includegraphics[width=0.9\columnwidth]{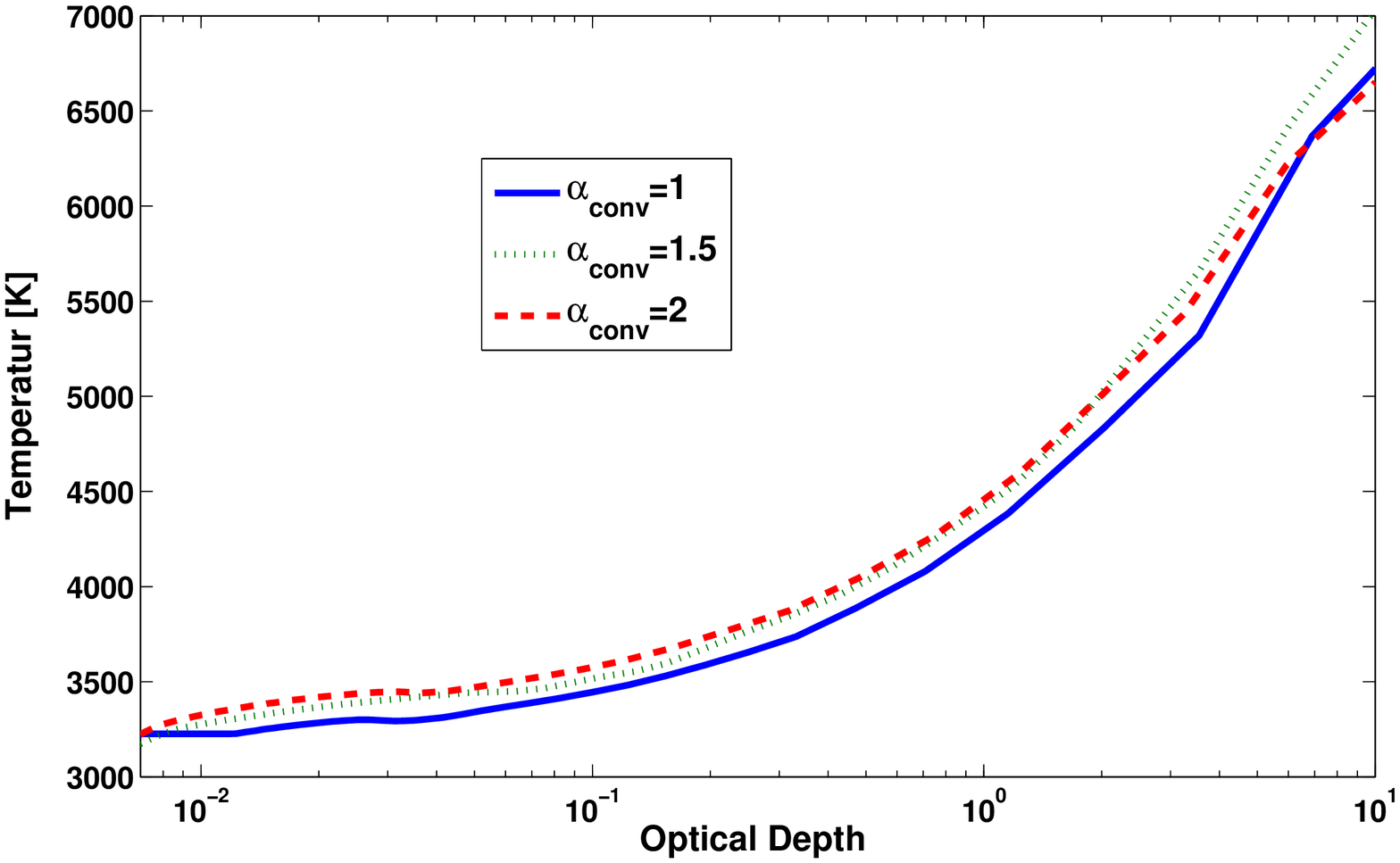}
 \caption{The vertical temperature structure for $\alpha_{ml}=1,1.5$ and $2$ (top) and as
 function of the optical depth (bottom).
 The model parameters are: $\alpha=0.03, M_{wd}=0.6,
 \Teff=5000$ and $\Rwd=75R_{wd}$}
 \label{fig:alpha_conv}
\end{figure}

As stated in the previous section we choose $\alpha_{ml}=1.5$ in order to compare
our results to those of HMDLH. In order to see how the disc structures and the emerging spectra are affected by varying
the mixing length parameter, we solved the equations assuming the two values delimiting the range allowed by solar type stars \citep{DG,gl}. Figure
\ref{fig:alpha_conv} shows the vertical structures of a $\Teff=5000$\,K  ring at $75\Rwd$ for $\alpha_{ml}=1,1.5$ and $2$. As $\alpha_{ml}$ decreases
convection becomes less efficient and the temperature therefore increases. Changing $\alpha_{ml}$ will affect the S-curves (discussed in the next
section). However, since changes in the upper part of the disc due to changes in $\alpha_{ml}$
are minor (Figure  \ref{fig:alpha_conv} bottom), the emerging spectrum is hardly affected by the assumed value of the mixing-length parameter.

\section{Full disc emission}
\label{sec:full_disc}

\subsection{Solutions at high accretion rate ($\alpha=0.3$)}
\label{subsec:hot}

\begin{figure}
 \vspace*{1em}
 \center
 \includegraphics[width=1.1\columnwidth]{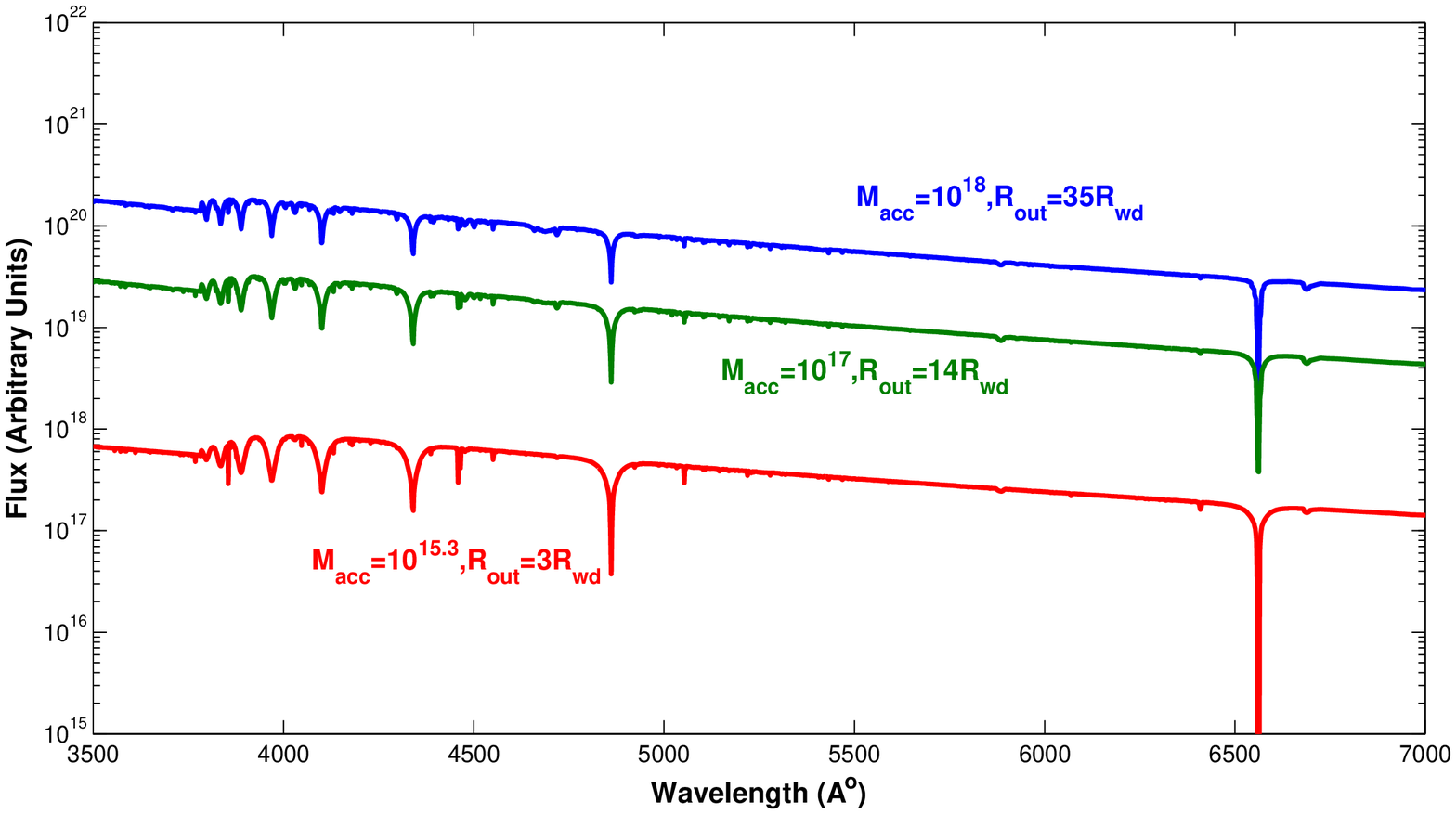}
 \includegraphics[width=1.1\columnwidth]{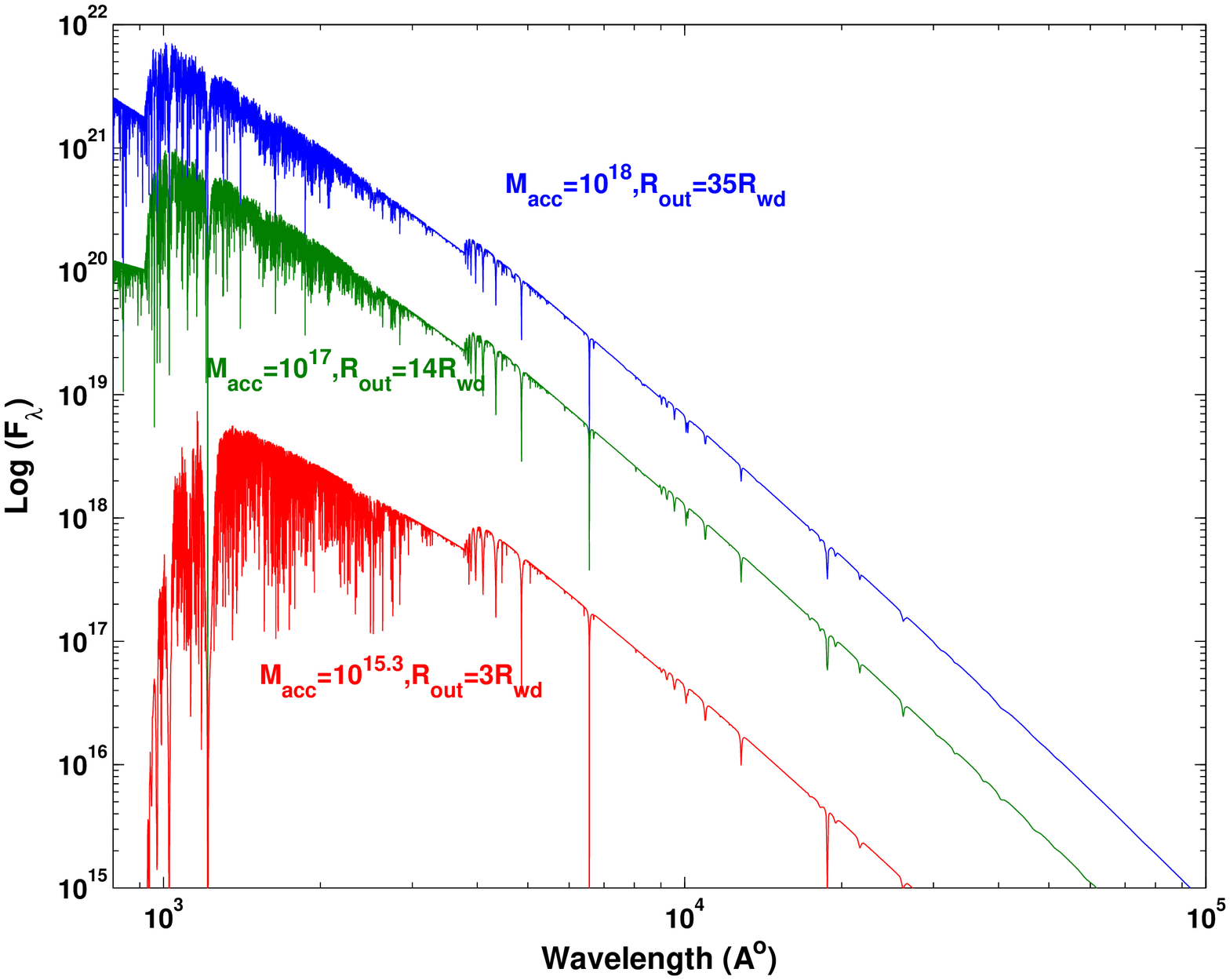}
  \caption{The full disc spectra for three values of accretion rates
  and outer radii corresponding to the decay from outburst.
  $\dot M=10^{18}\,, 10^{17}\,, 10^{15.3}\,\rm g\,s^{-1}$. The outer
  radius of the hot disc is respectively at $ 35\,, 14\, \rm and\, 3\, \Rwd$. ($\alpha=0.3, \Mwd=0.8\Msun$). }
 \label{fig:total_flux_03_18}
\end{figure}

\begin{figure}
 \vspace*{1em}
 \includegraphics[width=0.9\columnwidth]{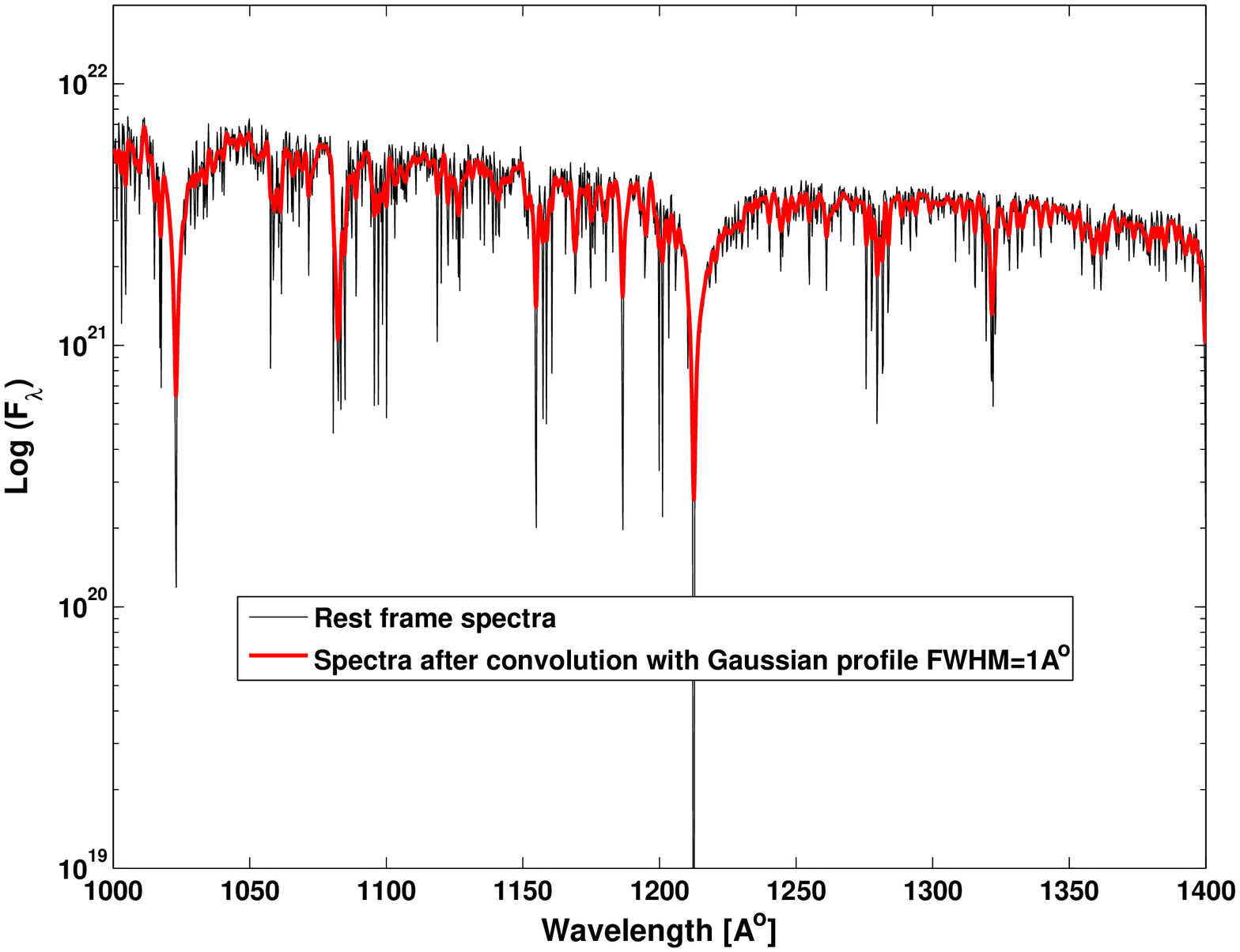}
  \caption{Rest frame spectra (black) and spectra shown after convolution (red) with a Gaussian instrumental profile
  (see text for details). The parameters are the same as in Fig. \ref{fig:total_flux_03_18} for $R_{\rm out}=35\Rwd$.}
 \label{fig:total_flux_03_18b}
\end{figure}
As recalled in Sect. \ref{subsec:dim}, during the dwarf-nova outbursts accretion discs decay from maximum through a sequence of hot configurations with decreasing accretion rate and outer radii \citep[see e.g. Fig. 5 in][]{hmdl98}. Spectra of three such configurations are shown in  Figure \ref{fig:total_flux_03_18} where $\alpha=0.3,
\Mwd=0.8\, \Msun$ have been assumed. The disc is seen face-on ($i=0^o$) and the white-dwarf spectrum has not been added.
The configurations correspond to three values of the accretion rate: $\dot M=10^{18}\,, 10^{17}\,, 10^{15.3}\,\rm g\,s^{-1}$
(maximum temperature $\Teff= 70\, 000\,, 24\, 000\, \rm and\, \rm 16\,000$K respectively). The outer radii correspond to the position of the cooling front, i.e. to the radius where $\Teff\approx T^+_{\rm crit}$. Here we chose 35, 14 and 3 $\Rwd$ ($2.45\times 10^{10}$, $9.8\times 10^{9}$ and $2.1\times 10^9$\, cm) .

In all these spectra one can
identify the Balmer lines (at $3970\,{\AA},4100\,{\AA},4340\,{\AA},4860\,{\AA}, 6560\,{\AA}$) as well as the Paschen series and the Lyman lines.
Until now we have presented unconvolved ``raw spectra" to show the direct output of the radiative transfer calculation.
Figure \ref{fig:total_flux_03_18b} shows the comparison between the rest- frame spectra of an accretion disc in the UV part of the wavelengths and the
same spectrum  convolved with a Gaussian instrumental profile having a FWHM of 1\AA \ sampled every 0.15\AA.

\subsection{Comparison with the \citet{wh98} accretion disc
spectra}
\label{subsec:wh}

\citet[][hereafter WH]{wh98} calculated with TLUSTDISK and
SYNSPEC a large grid of far- and mid-ultraviolet spectra
(850-2000 {\AA}) of the integrated light from steady-state
accretion discs in bright cataclysmic variables. These data
have being used extensively in the recent years for studying
the observed accretion disc spectra. In WH model the disc
spectra are calculated in four steps. In the first step they
calculate the vertical structure of concentric annuli, with
each annulus behaving as an independent plane-parallel
radiating slab. The energy balance is enforced between
radiative losses at the disc surface and heat generation due to viscosity. Once the vertical structure of the annuli is
obtained, the radiative transfer equation is solved to compute
the local, rest--frame spectrum for each ring of the disc. In
the third stage the rest--frame intensities are combined to
generate an integrated disc spectrum, and in the fourth stage
the monochromatic fluxes are convolved with a Gaussian
instrumental broadening function and then re-sampled uniformly
in wavelength. All the published WH models are for disc
atmospheres which are optically thick and are viewed from a
distance of $100 \rm pc$.

\citet{wh98} also studied the effect of different instrumental
resolutions (Gaussian FWHM = 0.1, 1.0 and 3.5 \AA) of the UV
spectrum. The FWHM=0.1\AA \ corresponds to the resolution of
the FUSE spectrometers whereas 3.5\AA \ corresponds to the HUT
spectrometer. According to \citet{wh98} the differences between the 0.1\AA \ \ and 1\AA \ \ were barely noticeable, and they
applied a FWHM of 1\AA  \ \ for their calculations. Having
confirmed their results we also have chosen instrumental broadening with  FWHM=1\AA.

\begin{figure}
 \vspace*{1em}
 \includegraphics[width=0.9\columnwidth]{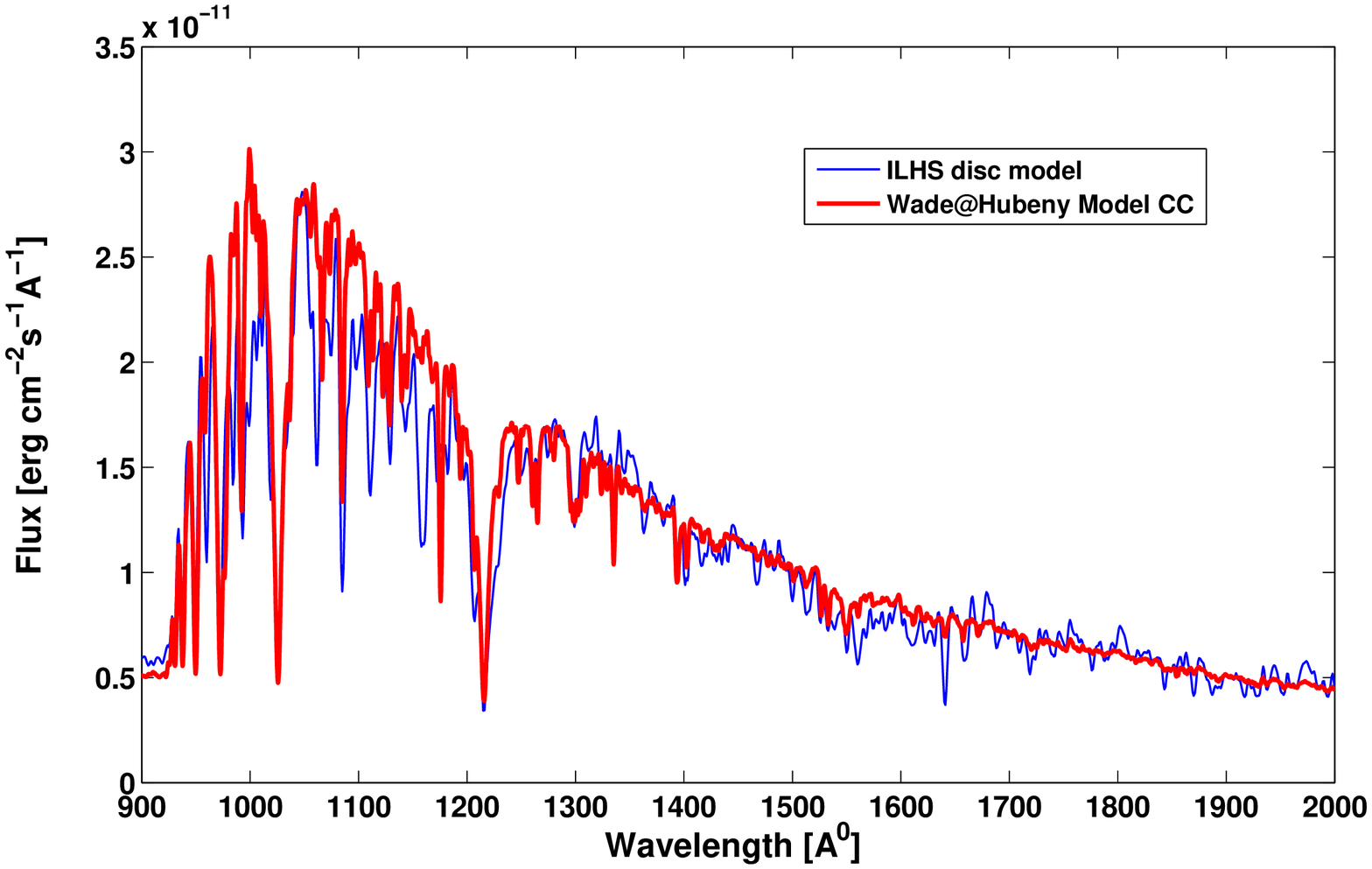}
 \includegraphics[width=0.9\columnwidth]{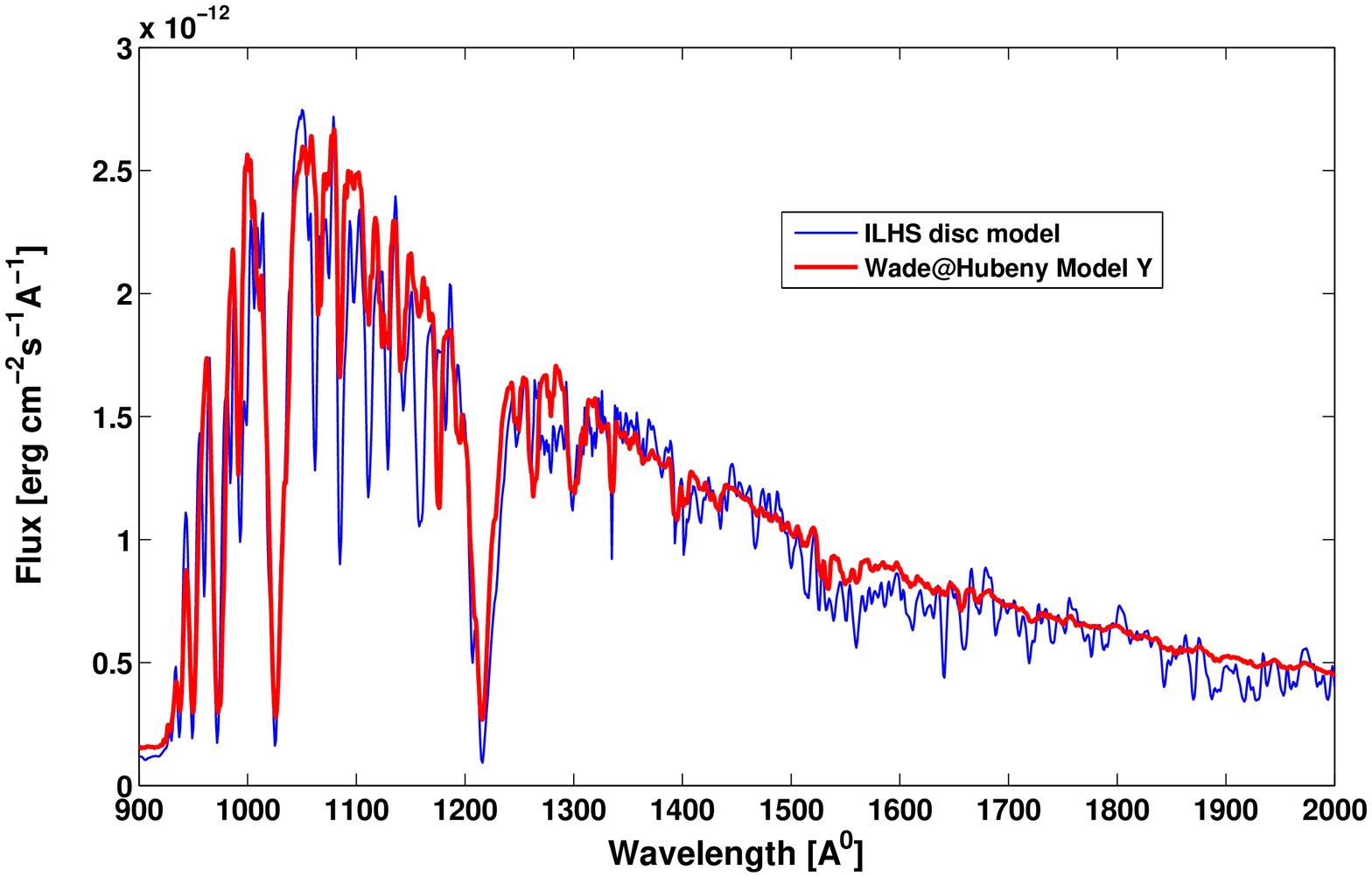}
 \caption{Comparison with spectra calculated by \citep{wh98}. Model cc (upper panel)
 and model y (lower panel) are compared with ILHS spectra calculated for
 $\alpha=0.3$ and the same $\Mwd$ and $\dot M$ as in \citet{wh98}. Both spectra
 are shown after convolution with a Gaussian instrumental profile.  }
 \label{fig:wadedc}
\end{figure}

Figure \ref{fig:wadedc} shows a comparison between the
{\citet{wh98}} accretion discs spectra (model cc and model y)
and the spectra obtained with the present ILHS programme. The
model parameters are: Model cc - $\Mwd=0.8 \Msun$ and
$\dot{M}=3.16\times 10^{-9} \Msun \rm yr^{-1}$ and for model y
- $\Mwd=1.21 \Msun$ and $\dot{M}=10^{-10 }\Msun \rm yr^{-1}$.
We used the same number of rings.

The difficulty with comparing the two schemes lies in the formulation of the viscosity prescription.
We are using the $\alpha$ prescription whereas
WH define the viscosity coefficient through the value of Reynolds number of the flow,
the vertical stratification being controlled by a parameter
$\zeta$ \citep{kh86}. The relation between the two parameterizations of disc viscosity is
rather complicated and non-local \citep[see][]{hh,nagel3}.
The relation between the viscosity parameter $\alpha$ and the Reynolds number $R_e$ of \citet{kh86} can be written as
\begin{equation}
\alpha=\left(\frac{v_K}{c_s}\right)^2\,R_e^{-1}\, ,
\label{eq:re}
\end{equation}
where $v_K$ is the Keplerian speed and $c_s$ the speed of sound.
The parameter $\zeta$ represents the vertical stratification of viscosity through:
\begin{equation}
\nu(z)=\left(\nu_0\frac{\varsigma}{\Sigma}\right)^{\zeta} ,
\label{eq:zeta}
\end{equation}
to be compared with the ansatz $\nu(z)=\alpha\,P(z)$ used in our scheme.

We estimated that our $\alpha=0.3$ should be roughly equivalent to the $Re=5000$ and $\zeta=2/3$ of WH but the correspondence between the two different descriptions of the viscosity mechanisms is not exact. Despite of that the two spectra are very similar and the small differences are in practice difficult to pinpoint. The comparison of the two methods presented in \citet{Idan2} showed a rather serious difference around 1300{\AA}: a clear absorption feature seen in \citet{wh98} was missing in ILHS. After a careful study of both the code and the OP database we arrived to the conclusion that there is an error in the opacity data (silicon lines misplaced). This has been confirmed and corrected by Delahaye (private communication). Spectra in Figure \ref{fig:wadedc} have been calculated with the corrected opacity tables. Now both spectra show an absorption feature at 1300{\AA}. The remaining differences most probably are due to the differences in the parametrization
(and stratification) of viscosity.


\section{Cold, quiescent dwarf-nova disc}
\label{sec:non-stat}

According to the disc instability model, the disc in  the
quiescent phase of the dwarf-nova outburst is characterized by
a low $\alpha$ and a low (quasi-constant) effective temperature. The effective
temperature in the entire disc must be lower than the critical
value \citep[see][]{l01}
\begin{equation}
T_{\rm eff}^{-} \approx 5800~\left(\frac{r}{10^{10}\rm cm}\right)^{-0.09}~\,\rm K
\label{eq:Teff_crit_low}
\end{equation}
and the critical midplane temperature in the quiescent state
must be lower than\footnote{Whereas the fit critical of $T_{\rm
eff}$ is accurate within few percent, the accuracy of the fit
to the critical central temperature is no better than $20\%$.
This explains the difference with the fits given in
\citet{ldk}, which in addition contain typing errors. The
formula in Eq. A1 of this paper should read: $T_{\rm c}^{-}
=8240~\alpha_{0.1}^{ -0.14}~R_{10}^{-0.08}~M_1^{ 0.04}\,\rm
K$.}:
\begin{equation}
T_{\rm c}^{-} \approx 9000~\alpha^{-0.13}~\left(\frac{r}{10^{10}\rm cm}\right)^{-0.01}~\,\rm K.
\label{tctau}
\end{equation}
Both observations and models show that for most of the
quiescence duration the effective temperature profile is roughly
parallel to the critical one. Therefore as examples of
spectra of quiescent dwarf-nova disc we calculated models
with constant (with radius) effective temperature. In
such non-equilibrium discs accretion rate is not constant but
increases strongly ($\sim r^{2.7}$) with radius.

Let us stress that low values of $\alpha$\, ($\lta 0.03$) in quiescence
are an essential ingredient of the DIM. This is because observations fix the value of the
hot disc $\alpha(\rm hot)$
to $\sim 0.2$ \citep{Smak99} and $\alpha(\rm hot)/\alpha(\rm cold)\gta 4 - 5$ is required since otherwise the model produces no dwarf-nova outbursts. Since according to the DIM the values of the critical surface densities (see Eq. \ref{eq:sigmamin}) are rather high and the values of $\alpha$ in quiescence must be rather low, the model requires quiescent discs to be optically thick, especially in their outer regions.

On this subject observations are at best ambiguous.
For example, according to \citet[][]{little01} \citep[confirmed by][]{cooperw} optical observations of the quiescent dwarf nova IP Peg seem to suggest that the outer part of the disc in this system is optically thin.
On the other hand \citet{ribeiro07} find that infrared observations imply an optically thick outer disc.
On the other hand practically all observed dwarf-novae in quiescence have flat radial-temperature ($< 5000$\,K) profiles and strongly
inward-decreasing accretion rate \citep{wetal86,wetal89,froning99} in excellent agreement with the prediction of the DIM. The presence of strong emission
lines is not an evidence for the optical thinness of the disc \citep{Smak91} since they can be produced in a chromosphere above the disc. Also, as suggested by \citet{Vrielmann}, the quiescent disc could be clumpy, with optically thin patches
interspersed among optically thick gruels. In addition IP Peg shows strong emission lines also in outburst \citep{marshorne90}.
The whole subject is therefore rather controversial \citep[``based on current knowledge of such discs,
optically thick regions might or might not be present", according to][]{HRB05} so that it is clearly very important to be able to model correctly and consistently the emission spectra of quiescent dwarf novae if one wishes to solve this fundamental problem.

Here we present the first spectra calculated for  self-consistent cold disc structures with $\alpha=0.03, \Mwd=0.6\,M_{\odot}$ and a constant effective
temperature of 5000 and 3500K \citep[The second value corresponds to IP Peg in quiescence according to][]{ribeiro07}. The outer radius in the model was assumed to be $75 \Rwd$.

\begin{figure}
 \vspace*{1em}
 \includegraphics[width=0.9\columnwidth]{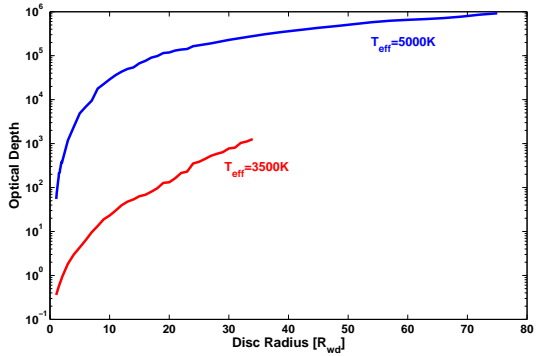} \\
 \caption{The optical depth as function of the disc radii for disc models with $\alpha=0.03, \Mwd=0.6\Msun$ and two values of effective temperatures \Teff=5000\,K, and \Teff=3500\,K. Above 35 $\Rwd$ the 3500 K disc is fully convective and its spectrum cannot be calculated by our method.}
 \label{fig:opt_depth}
\end{figure}

\begin{figure}
 \vspace*{1em}
 \center
 \includegraphics[width=0.9\columnwidth]{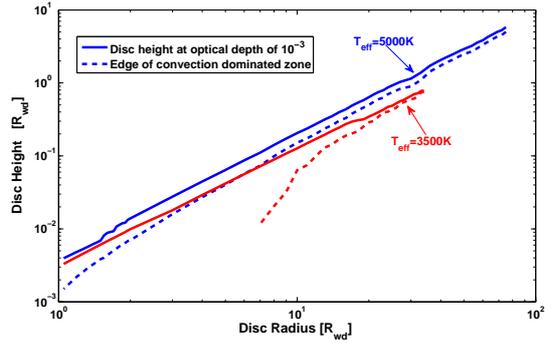} \\
 \caption{The disc height at optical depth of $10^{-3}$ (solid line) and the height of the convection zone (dashed line)
   for two values of effective temperatures \Teff=5000\,K, and \Teff=3500\,K; ($\alpha=0.03,\Mwd=0.6\Msun$).}
 \label{fig:edge_convection}
\end{figure}

The 5000 K quiescent disc is everywhere optically
thick (Figure \ref{fig:opt_depth}); the optical depth in the outer regions reaching $10^6$. The disc is everywhere convection dominated.  As shown above, the convective flux determines the vertical structure of cold accretion discs and obviously cannot be ignored through the part of the disc
where the optical depth is larger than 5-10. The 3500 K disc is optically thin in its innermost regions but
starting from $\sim 10\Rwd$ the optical depth $\tau > 10$ and reaches $10^4$ at the outer edge.
Figure \ref{fig:edge_convection} shows the extension of the convective zones for cold quiescent discs.
The height of the disc corresponding to optical depth of $10^{-3}$ is marked with a
solid line while the height of the convection zone (where the convective flux is equal to half of the viscous flux) is marked with a dashed line.
As seen in this figure the depth of the convection-dominated zone increases with radius. For the 5000 K disc it almost reaches the base of the photosphere for $R\sim 75 \Rwd$.
The inner regions ($R\lta 7\,\Rwd$) of the 3500 K disc are radiative but at 35 $\Rwd$ the convection dominated zone reaches the photosphere. For larger disc radii the "leftover" radiative flux is too minuscule to be sensibly included into the radiative transfer code.
In any case assuming in the framework of the DIM the optical thinness of the outer disc is totally \textsl{ad hoc} and does not provide any understanding of the quiescence according to this model.

\begin{figure}
 \vspace*{1em}
 \includegraphics[width=0.9\columnwidth,height=4.6truecm]{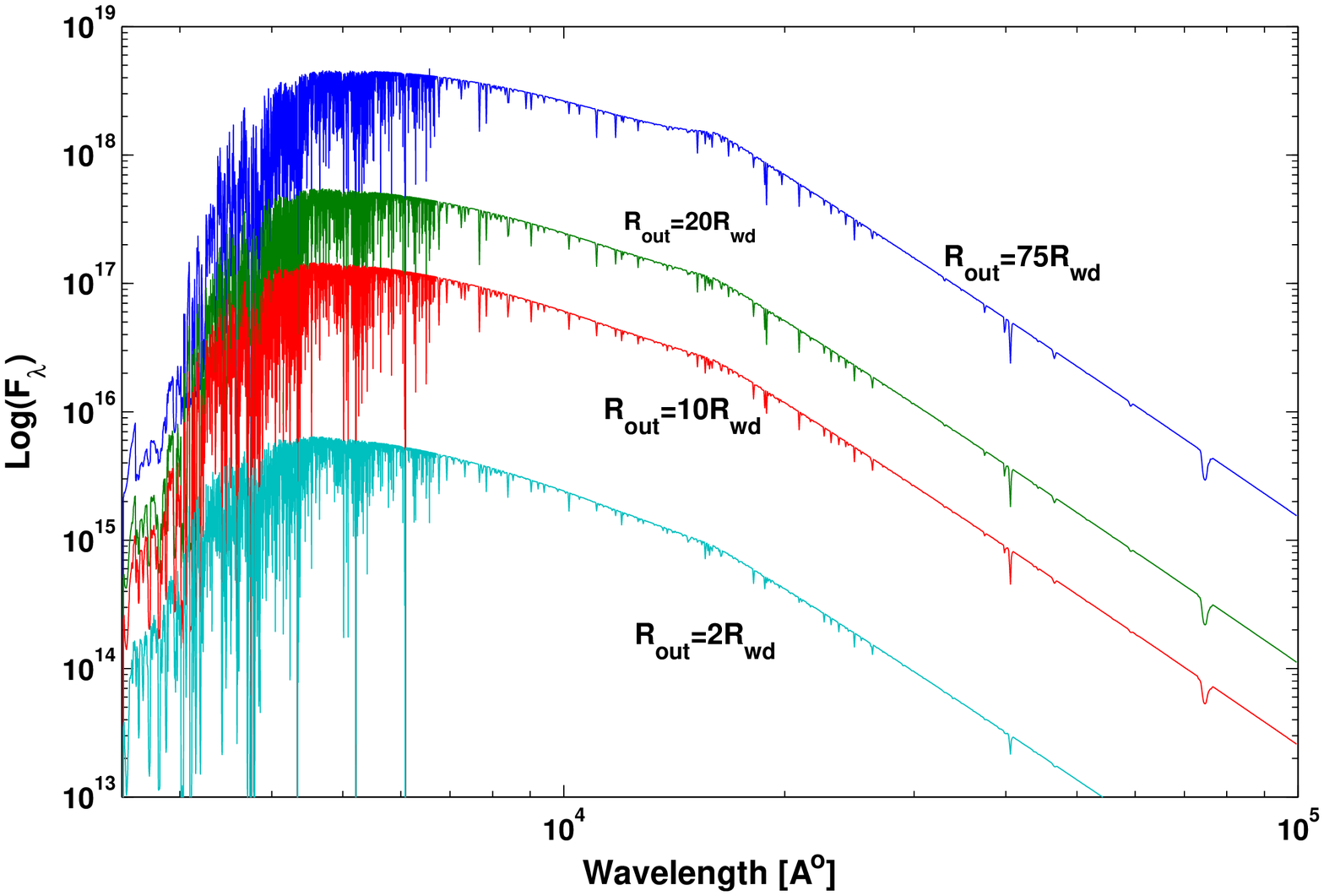}
 \includegraphics[width=0.9\columnwidth]{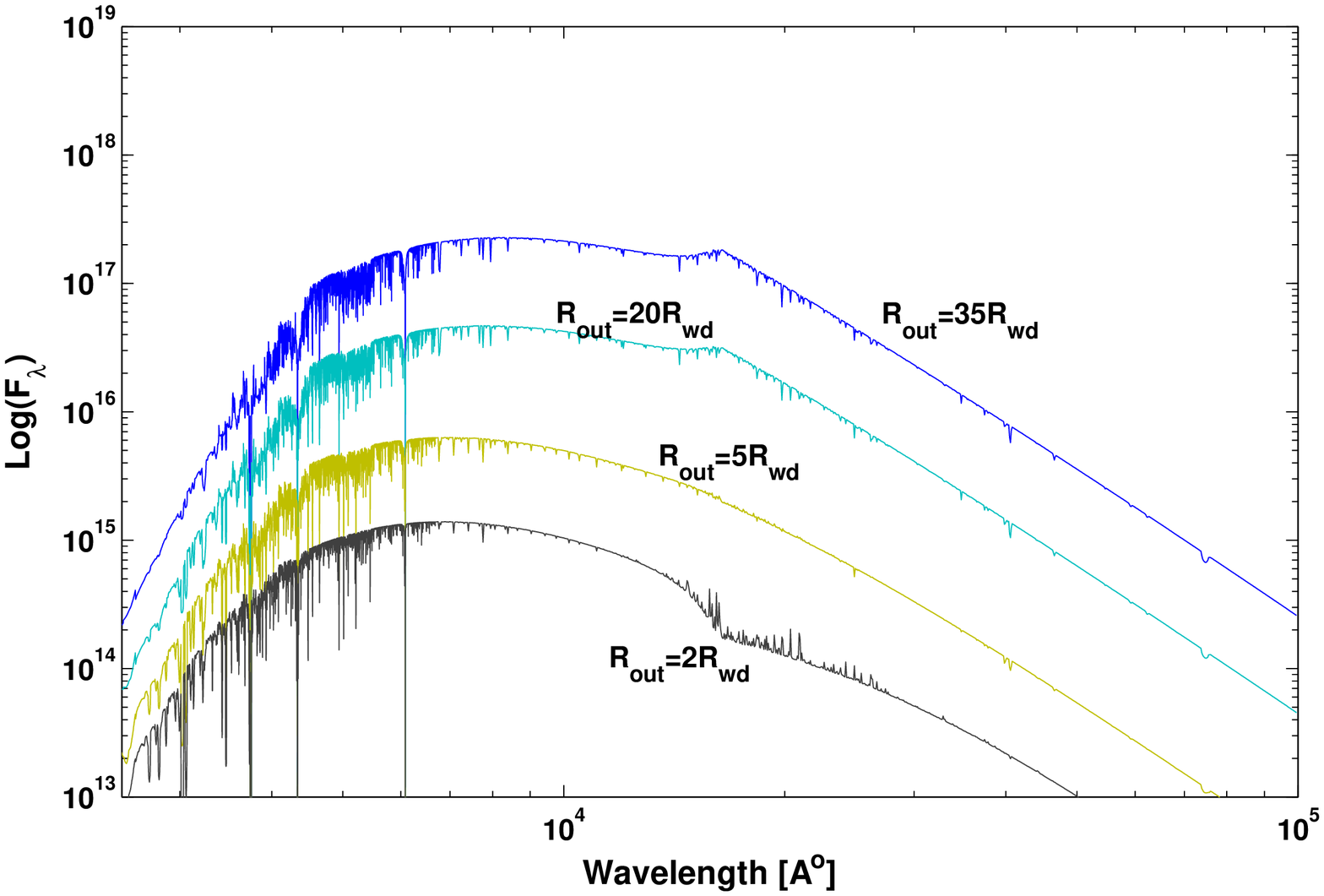}
 \\
 \caption{The integrated disc spectrum as function of the outer disc radius for
 cold discs with $\Teff=5000$\,K (upper panel) and $\Teff=3500$\,K (lower panel); $\alpha=0.03, \Mwd=0.6\Msun$, $\Rwd= 8.5\times 10^8\,$cm. }
 \label{fig:total_flux_003_5000}
\end{figure}

\begin{figure}
 \vspace*{1em}
 \includegraphics[width=0.9\columnwidth]{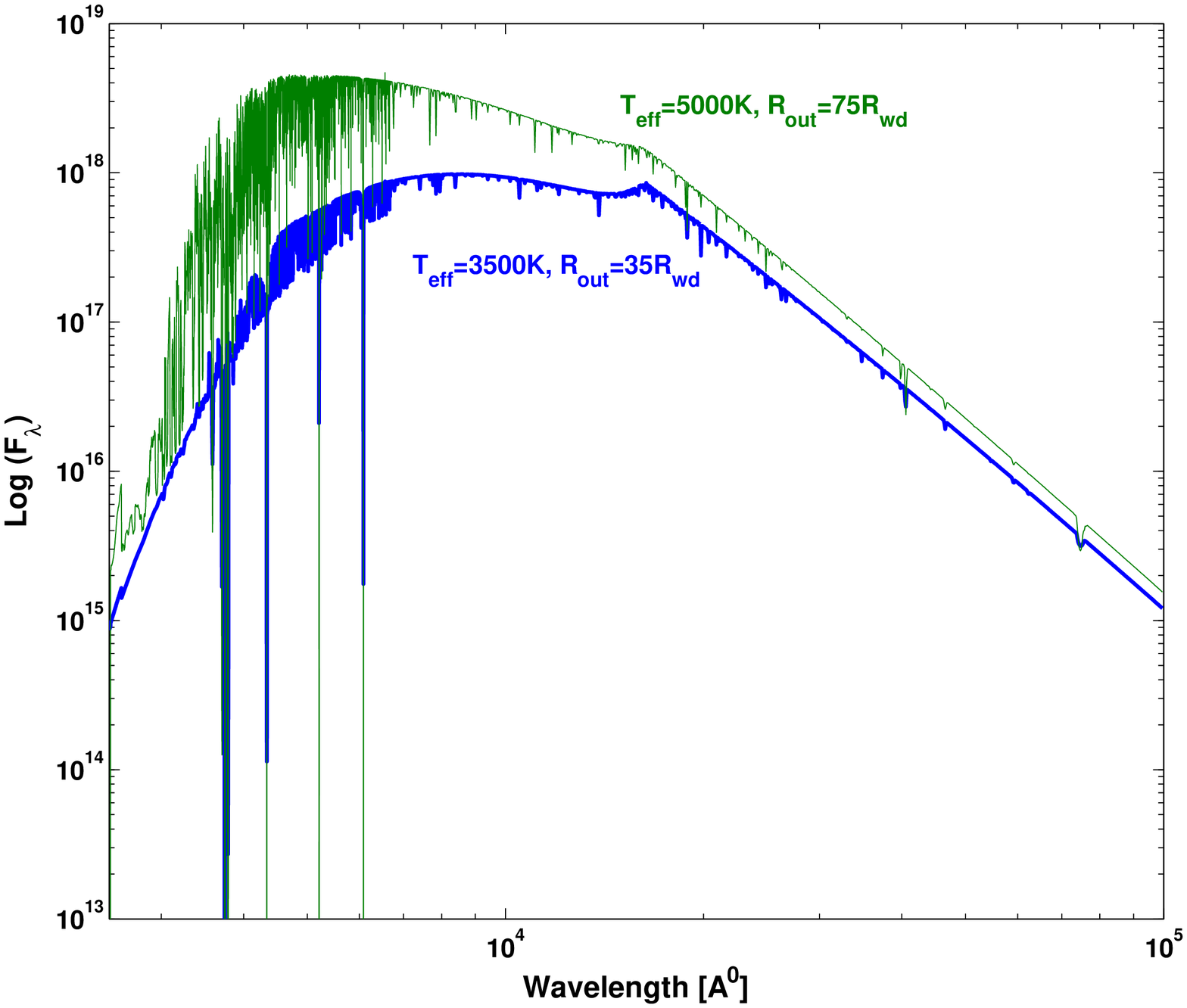} \\
 \caption{The integrated disc spectrum for two low values
 of $\Teff=5000$ and 3500 ($\Mwd=0.6\Msun$, $\alpha=0.03$).}
 \label{fig:spect_3500_5000}
\end{figure}

The total fluxes obtained for cold discs with an outer radius of $75\,\Rwd$ are shown in Fig. \ref{fig:total_flux_003_5000} and \ref{fig:spect_3500_5000}. The spectra in the range of $3000-10000$\,{\AA} are nearly flat as observed and no Balmer jump is present. As expected there is no significant contribution from the disc to the UV band. The 3500 K shows a Brackett jump in absorption, except for case when radiation comes from the innermost disc regions only; then the jump is in emission and weak emission lines are present in its vicinity. This reflects obviously the optical thinness of these regions but is unlikely to have observable consequences.

As mentioned above quiescent dwarf nova spectra often show strong and broad hydrogen and helium emission lines. The spectrum supposed to represent a
cold dwarf-nova disc in Figure \ref{fig:total_flux_003_5000} has only absorption line. The
temperature in the photosphere of the cold disc are too low to
produce emission lines. \citep[As mentioned above, the emission lines we reported in][were an artefact of insufficient
resolution of the computational grid]{Idan2}.

For low $\alpha$s we were unable to produce a ``corona" by the \citet{sw86} instability. Apparently heating was always compensated by cooling even
high in the disc atmosphere.
Therefore it seems that to obtain emission lines one should take
into account the effect of irradiation by the white dwarf
\citep[as suggested by][]{Smak91}.
\citet{nagel3} have included disc irradiation in this
context. They calculated spectra of an accretion disc in
quiescence and in outburst attempting to reproduce observations of the dwarf-nova SS Cyg. Their model reproduces the hydrogen Balmer emission lines but this interesting result was
obtained by {\sl assuming} that the disc outer ring is optically thin. As reminded above
such an assumption is in contradiction with the DIM results and, as well
known, implies a value of the viscosity parameter $\alpha > 1$. Indeed, \citet{nagel3} assume in the outer ring a value of $R_e$ ten time lower than in
the rest of the disc. As shown by \citet{Idan1}, in a quiescent disc the requirement of optical thinness inevitably leads to high values of the
viscosity parameter. Because of these
arbitrary choices it is unclear what is the meaning of their results.

\section{Conclusions}
\label{sec:concl}

We have constructed a numerical code solving radiative transfer in accretion discs for a range of effective temperatures reaching values
as low as 3000\,K. As far as we know, our code is the only one
taking into account convective energy transport, including the cases when convection is dominating the energy transport.
At high temperatures, in the case of fully radiative accretion discs, our spectra are in a very satisfactory agreement with the calculations of \citet{wh98} despite the different viscosity prescriptions respectively used. The $\alpha$-viscosity prescription of our code is
used in the Disc Instability Model and in general to characterize the structure of accretion discs.

Although our code includes convective energy transport it fails to account for important effects that have been successfully taken into account by other authors. For example we assume LTE, whereas the AcDC code \citep{nagel04} describes the NLTE structures. These different approaches
are therefore only milestones on the road towards a complete description of accretion disc emission. Unfortunately the main obstacle one
will have to overcome is the persisting lack of a reliable description of the disc's ``viscosity" stratification.

Our code has been designed to work in synergy with the DIM code of HMDLH. This goal has been achieved for most of the range of the relevant parameters. We find, in agreement with the predictions of the DIM, that quiescent dwarf-nova discs must be optically thick. In absence of spontaneously formed disc corona emission lines observed in quiescent dwarf-novae have to be the result of irradiation by the white dwarf (and the boundary layer) of an optically thick disc as suggested by \citet{Smak91}.

In a subsequent article we will include the effects of limb darkening
and produce a grid of disc spectra to be compared with
observations of various types of cataclysmic variables and various phases of the dwarf-nova outburst cycle. In
particular we will focus on the quiescence discs and the effect of the irradiation by the white dwarf and the formation
of emission lines. Combining ILHS with the time dependent HMDLH code will provide a tool for the study of dwarf nova outburst cycles.

In particular it will be possible to compare observed quiescent dwarf nova spectra with the model spectra of cold non-equilibrium accretion discs and hot white dwarfs, instead of assuming \citep[as in e.g.][and references therein]{urbsion} that the disc in such a state is hot and in equilibrium.

The structure of our code allows easy inclusion of various tabulated opacity data. In a future work we plan calculating spectra of helium discs
and very cold discs in which molecular opacities are important.

\acknowledgements{} We thank Franck Delahaye and Claude Zeippen  for their valuable help with the OP opacities. We will miss the enlightening discussions with our late friend Reiner Wehrse  and his always valuable suggestions. Discussions with Ivan Hubeny have been of great help. Some remarks by the referee of the first submitted version of this paper were also helpful. We thank the second referee for very constructive and useful comments. JPL acknowledges support from the French Space Agency CNES and from Polish Ministry of Science and Higher Education within the project N N203 380336.

\end{document}